\documentclass[12pt]{article}
\usepackage{epsfig}
\usepackage{amssymb}
\usepackage{amsmath}
\usepackage{amsfonts}
\usepackage{graphicx}
\usepackage{mathrsfs}
\usepackage[dvips]{color}
\usepackage{multirow}

\newcommand{\bsigma}{\boldsymbol{\sigma}}

\newcommand{\bnabla}{\boldsymbol{\nabla}}

\newcommand{\R}{\mathbb{R}}
\newcommand{\C}{\mathbb{C}}

\newcommand{\ft}{\mathfrak{t}}

\newcommand{\fz}{\mathfrak{z}}

\newcommand{\fK}{\mathfrak{K}}

\newcommand{\fS}{\mathfrak{S}}

\newcommand{\bba}{\mathbf{a}}

\newcommand{\bbe}{\mathbf{e}}

\newcommand{\bh}{\mathbf{h}}
\newcommand{\bk}{\mathbf{k}}

\newcommand{\bbr}{\mathbf{r}}

\newcommand{\bv}{\mathbf{v}}

\newcommand{\bzero}{\mathbf{0}}
\newcommand{\bA}{\mathbf{A}}

\newcommand{\bB}{\mathbf{B}}
\newcommand{\bC}{\mathbf{C}}

\newcommand{\bF}{\mathbf{F}}
\newcommand{\bH}{\mathbf{H}}
\newcommand{\bI}{\mathbf{I}}
\newcommand{\bJ}{\mathbf{J}}
\newcommand{\bL}{\mathbf{L}}
\newcommand{\bK}{\mathbf{K}}

\newcommand{\bM}{\mathbf{M}}

\newcommand{\bS}{\mathbf{S}}
\newcommand{\bT}{\mathbf{T}}
\newcommand{\bX}{\mathbf{X}}
\newcommand{\bY}{\mathbf{Y}}
\newcommand{\bU}{\mathbf{U}}

\newcommand{\cH}{\mathcal{H}}
\newcommand{\cE}{\mathcal{E}}
\newcommand{\cF}{\mathcal{F}}

\newcommand{\cL}{\mathcal{L}}
\newcommand{\cM}{\mathcal{M}}

\newcommand{\cS}{\mathcal{S}}

\newcommand{\be}{\begin{equation}}
\newcommand{\ee}{\end{equation}}
\newcommand{\bea}{\begin{eqnarray}}
\newcommand{\eea}{\end{eqnarray}}
\newcommand{\nn}{\nonumber}
\newcommand{\kt}{\rangle}
\newcommand{\br}{\langle}

\newcommand{\ed}{\end{document}}

\newcommand{\bi}{\begin{itemize}}
\newcommand{\ei}{\end{itemize}}

\newcommand{\bce}{\begin{center}}
\newcommand{\ece}{\end{center}}

\newcommand{\sD}{\mathscr{D}}

\newcommand{\sF}{\mathscr{F}}

\newcommand{\sH}{\mathscr{H}}

\newcommand{\sT}{\mathscr{T}}
\newcommand{\sU}{\mathscr{U}}
\newcommand{\RE}{{\rm Re}}
\newcommand{\IM}{{\rm Im}}

\newcommand{\bcE}{{\boldsymbol{\cE}}}
\newcommand{\bcH}{{\boldsymbol{\cH}}}
\newcommand{\brE}{{\boldsymbol{\rm E}}}
\newcommand{\brH}{{\boldsymbol{\rm H}}}
\newcommand{\bphi}{{\boldsymbol{\phi}}}
\newcommand{\bPhi}{{\boldsymbol{\Phi}}}

\newcommand{\bPi}{{\boldsymbol{\Pi}}}

\newcommand{\bEpsilon}{\mbox{\Large${\boldsymbol{\epsilon}}$}}
\newcommand{\bxi}{{\boldsymbol{\xi}}}
\newcommand{\bXi}{{\boldsymbol{\Xi}}}

\newcommand{\bup}{{\boldsymbol{\Upsilon}}}
\newcommand{\bfeta}{{\boldsymbol{\eta}}}
\newcommand{\bfTheta}{{\boldsymbol{\Theta}}}
\newcommand{\bfS}{{\boldsymbol{\fS}}}

\begin{document}

\title{Transfer-matrix formulation of the scattering of
electromagnetic waves and broadband\\ invisibility in three
dimensions}

\author{Farhang Loran\thanks{E-mail address: loran@iut.ac.ir}
~and Ali~Mostafazadeh\thanks{E-mail address:
amostafazadeh@ku.edu.tr}\\[6pt]
$^*$Department of Physics, Isfahan University of Technology, \\ Isfahan 84156-83111, Iran\\[6pt]
$^\dagger$Departments of Mathematics and Physics, Ko\c{c} University,\\  34450 Sar{\i}yer,
Istanbul, Turkey}

\date{ }
\maketitle

\begin{abstract}

We develop a transfer-matrix formulation of the scattering of
electromagnetic waves by a general isotropic medium which makes use
of a notion of electromagnetic  transfer matrix $\widehat{\bM}$ that
does not involve slicing of the scattering medium or discretization
of some of the position- or momentum-space variables. This is a
linear operator that we can express as a $4\times 4$ matrix with
operator entries and identify with the S-matrix of an effective
nonunitary quantum system. We use this observation to establish the
composition property of $\widehat{\bM}$, obtain an exact solution of
the scattering problem for a non-magnetic point scatterer that
avoids the divergences of the Green's function approaches, and prove
a general invisibility theorem. The latter allows for an explicit
characterization of a class of isotropic media $\cM$ displaying
perfect broadband invisibility for electromagnetic waves of
arbitrary polarization provided that their wavenumber $k$ does not
exceed a preassigned critical value $\alpha$, i.e., $\cM$ behaves
exactly like vacuum for $k\leq\alpha$. Generalizing this phenomenon,
we introduce and study $\alpha$-equivalent media that, by
definition, have identical scattering features for $k\leq\alpha$.
\vspace{2mm}

%\noindent PACS numbers: 03.65.Nk, 42.25.Bs\vspace{2mm}

%\noindent Keywords:

\end{abstract}

\section{Introduction}
\label{S1}

Scattering of electromagnetic (EM) waves has been a subject of
intensive research for over a century. There are excellent
monographs covering its various aspects
\cite{newton,TKD,born-wolf,chew}. Most of the theoretical
developments in this subject are based on the use of Green's
functions and the related integral equations \cite{born-wolf,VCL}.
The purpose of the present article is to offer an alternative
formulation of the scattering of EM waves by isotropic
media\footnote{By an isotropic medium we mean a linear stationary
medium whose electromagnetic properties are characterized by scalar
permittivity $\varepsilon$ and permeability profiles $\mu$,
\cite{jackson}. These are generally complex scalar functions of
space.} that has significant advantages over the standard Green's
function methods. This formulation relies on a notion of transfer
matrix (operator) that can be employed for the study of the
scattering of EM waves by an arbitrary isotropic medium.

The use of transfer matrices in the study of the scattering of waves has a long history. There are hundreds of research publications on theoretical aspects and applications of transfer matrix theories. The notion of a transfer matrix  was initially developed for solving one-dimensional scattering problems for scalar waves \cite{jones-1941,abeles,thompson} and found important applications in the study of (effectively) one-dimensional optical \cite{teitler-1970,berreman-1972,yeh,schubert-1996,katsidis-2002,Hao-2008}, condensed matter \cite{abrahams-1980,ardos-1982,pendry-1982,sheng-1996,wortmann-2002,li-2009,zhan-2013}, and acoustic systems \cite{levesque,hosten,wang-2001}. The multichannel extensions of the transfer matrix have also been considered \cite{pereyray-1998a,pereyray-2002,pereyray-2005,Shukla-2005,anzaldo-meneses-2007} and its generalization to two- and three-dimensional systems were developed through appropriate discretizations of the transverse degrees of freedom to the scattering axis \cite{pendry-1984,pendry-1990a, pendry-1990b,pendry-1994,mclean,ward-1996,pendry-1996}. The common feature of the transfer matrices considered in the above references is that they are numerical matrices (of different sizes) storing the information about the scattering properties of the system and fulfilling an extremely useful composition rule. The latter allows for the calculation of the scattering properties of a medium by slicing it into thin layers, obtaining the transfer matrix for each slice, and determining the transfer matrix for the medium from those of its slices by invoking their composition rule.

The simplest example of a transfer matrix is that of time-independent scattering theory in one dimension. Consider a possibly complex-valued scattering potential $v(x)$ and time-harmonic scalar waves $e^{-i\omega t}\psi(x)$ satisfying the Schr\"odinger equation:
    \be
    -\psi''(x)+v(x)\psi(x)=k^2\psi(x).
    \label{sch-eq-1d}
    \ee
Here $k$ is the wavenumber for the incident wave which takes real and positive values. Suppose that $v(x)$ decays to zero for $x\to\pm\infty$ at such a rate that every solution of  (\ref{sch-eq-1d}) fulfills the asymptotic boundary conditions:
    \be
    \psi(x)\to A_\pm e^{ikx}+B_\pm e^{-ikx}~~{\rm for}~~x\to\pm\infty,
    \label{symp-1d}
    \ee
where $A_\pm$ and $B_\pm$ are possibly $k$-dependent complex coefficients. The transfer matrix of $v(x)$ is the unique $2\times 2$ matrix $\bM$ that satisfies
    \be
    \left[\begin{array}{c}
    A_+\\ B_+\end{array}\right]=\bM\left[\begin{array}{c}
    A_-\\ B_-\end{array}\right],
    \label{M=1D}
    \ee
and is independent of $A_-$ and $B_-$, \cite{razavy,sanchez}. This matrix has two important properties \cite{bookchapter}:
        \begin{enumerate}
        \item Its entries determine the reflection and transmission amplitudes of $v(x)$.
    \item If $v_1(x)$ and $v_2(x)$ are potentials such that $v(x)=v_1(x)+v_2(x)$,
        and the support of $v_1(x)$ is to the left of the support of $v_2(x)$, i.e.,    there is a real number $a$ such that $v_1(x)=0$ for $x>a$ and $v_2(x)=0$ for $x<a$, then the transfer matrices $\bM$, $\bM_1$, and $\bM_2$ of $v(x)$, $v_1(x)$, and $v_2(x)$ satisfy $\bM=\bM_2\bM_1$.
    \end{enumerate}
Property 1 shows that the computation of $\bM$ is equivalent to the solution of the scattering problem for $v(x)$. Property 2 allows for dissecting the support of $v(x)$ into a collection of intervals $I_j$, with $j=1,2,\cdots n$ and $I_j$ lying to the left of $I_{j+1}$, and reducing the scattering problem for $v(x)$ to that of its restriction onto $I_j$. Specifically, denoting the latter by $v_j(x)$, so that
    \begin{align*}
    &v_j(x):=\left\{\begin{array}{cc}
    v(x) & {\rm for}~x\in I_j,\\
    0 &{\rm otherwise,}\end{array}\right.
    &&\sum_{j=1}^n v_j(x)=v(x),
    \end{align*}
and labeling the transfer matrix of $v_j(x)$ by $\bM_j$, we have
    \be
    \bM=\bM_n\bM_{n-1}\cdots\bM_1.
    \label{composition}
    \ee
This is the main reason for the popularity of transfer matrices in dealing with scattering of scalar waves by effectively one-dimensional multilayer and locally periodic media \cite{abeles,yeh,pereyra,griffiths,yeh-book}.

The composition rule (\ref{composition}) has provided the main guideline for various generalizations of the transfer matrix that are tailored for dealing with the scattering of scalar and electromagnetic waves in higher dimensions \cite{pendry-1982,pendry-1990a,pendry-1994,mclean}. The main strategy leading to these generalizations is most succinctly summarized by McLean and Pendry \cite{mclean}: ``The methodology underlying all transfer matrix theories is extremely simple. The system that we wish to study is nominally partitioned into smaller sub-units (conventionally planar slices for a three-dimensional system). The relevant quantities that concern our theory are defined locally within the sub-units and the transfer matrix then links spatially adjacent units . The physical process of adding the subunits together sequentially to reproduce the bulk system is then described mathematically by the product of the individual transfer matrices for each sub-unit taken in the correct order.'' As we noted above, the definition of these generalized transfer matrices involves certain discretization of the position, momentum, or a mixture of these spaces. This leads typically to large numerical transfer matrices that require appropriate numerical treatments \cite{pendry-1990a,pendry-1994,mclean}.

In Ref.~\cite{pra-2016} we introduce a multi-dimensional transfer matrix whose definition does not rely on a slicing or discretization scheme. It rather makes use of a remarkable feature of the transfer matrix of the one-dimensional scattering theory (\ref{M=1D}), namely that it can be written as the S-matrix of a nonunitary effective two-level quantum system \cite{pra-2014a}. This means that we can express it as the time-ordered exponential of a non-Hermitian $2\times 2$ interaction-picture effective Hamiltonian $\sH(x)$;
    \be
    \bM=\sT~\exp\left\{-i\int_{-\infty}^\infty dx\: \sH(x)\right\},
    \label{M=exp}
    \ee
where $x$ plays the role of time. In other words, if we denote the evolution operator for the Hamiltonian $\sH(x)$ by $\sU(x,x_0)$, we have $\bM=\sU(\infty,-\infty)$.

Equation~(\ref{M=exp}) has many interesting implications for potential scattering in one dimension \cite{ap-2014,jpa-2014a}. More importantly, it provides an invaluable route towards a fundamental transfer-matrix formulation of the scattering of scalar waves in two and three dimensions \cite{pra-2016}. In this formulation the transfer matrix is given by the time-ordered exponential of a non-Hermitian effective Hamiltonian operator that acts in a certain infinite-dimensional function space. Therefore, the transfer matrix is no longer a numerical matrix; it is a $2\times 2$ matrix with operator entries. This formulation of potential scattering leads to a multidimensinal extension of the notion of unidirectional invisibility \cite{prsa-2016,p139}, allows for the exact solution of the scattering problem for infinite classes of scattering potentials in two and three dimensions \cite{pra-2017,jpa-2018}, and yields a method for constructing potentials that display perfect broadband invisibility in two dimensions \cite{ol-2017}. The latter model the scattering of transverse electric (TE) and transverse magnetic (TM) waves by certain effectively two-dimensional isotropic media.  Motivated by these developments, in this article, we propose a fundamental transfer-matrix formulation of the scattering of EM waves by general isotropic media. Unlike the EM transfer matrices considered in the literature \cite{pendry-1990a,pendry-1994,mclean,ward-1996,pendry-1996}, neither the definition nor the application of our EM transfer matrix requires a slicing of the scattering medium or a discretization of the position or momentum space.

The organization of the article is as follows. In Section~\ref{S2} we review the standard setup for EM scattering. In Section~\ref{S3} we provide the basic ingredients of our approach, introduce the EM transfer matrix, and explain how it can be used to determine the scattering amplitude and differential cross section for the scattering of EM waves by an arbitrary isotropic scattering medium. In Section \ref{S5}, we identify the transfer matrix with the S-matrix of an effective nonunitary quantum system and establish its composition property. In Section~\ref{S-point} we use our EM transfer matrix to solve the scattering problem for a nonmagnetic delta-function point scatterer. This turns out to avoid the divergences arising in the application of the Green's function methods to this problem and produces a finite expression for the scattering amplitude. The latter agrees with the known result provided that we identify the coupling constant of our approach with a renormalized coupling constant of the Green's function approach.  In Section~\ref{S6}, we discuss the application of EM transfer matrix in obtaining  a general criterion for perfect broadband invisibility that applies for isotropic media with no particular symmety. This allows us to construct an infinite class of isotropic media specified by a wavenumber scale $\alpha$ that do not scatter monochromatic EM waves with wavenumber $k\leq\alpha$ (and their superpositions.) In Section~\ref{S7}, we consider the problem of identifying isotropic media with identical scattering features for wavenumbers $k\leq\alpha$. We call these ``$\alpha$-equivalent" and give simple criteria for $\alpha$-equivalence. Section~\ref{S8} includes our concluding remarks.

\section{Basic setup for scattering of EM waves}
\label{S2}

Consider a time-harmonic EM wave with electric and magnetic fields, $e^{-i\omega t}\brE(\bbr)$ and $e^{-i\omega t}\brH(\bbr)$, that propagates in a stationary isotropic medium $\cM$ specified by the permittivity and permeability profiles: $\varepsilon(\bbr)$ and $\mu(\bbr)$. Here $\bbr:=x\,\hat\bbe_x+y\,\hat\bbe_y+z\,\hat\bbe_z$ is the position vector, $x$, $y$, and $z$ are coordinates in a Cartesian coordinate system with axes aligned along the unit vectors $\hat\bbe_x$, $\hat\bbe_y$, and $\hat\bbe_z$.

Let $\varepsilon_0$ and $\mu_0$ respectively denote the permittivity and permeability of the vacuum, and introduce the scaled electric and magnetic fields:
    \begin{align}
    &\bcE:=\sqrt{\varepsilon_0}\;\brE,
    &&\bcH:=\sqrt{\mu_0}\;\brH,
    \end{align}
and the relative permittivity and permeability:
    \begin{align*}
    &\hat\varepsilon:=\varepsilon/\varepsilon_0,
    &&\hat\mu:=\mu/\mu_0,
    \end{align*}
which are generally complex-valued functions of $\bbr$. Then the dynamical Maxwell equations take the form:
    \bea
    \bnabla\times\bcE-ik\hat\mu\,\bcH&=&0,
    \label{mx1}\\
    \bnabla\times\bcH+ik\hat\varepsilon\,\bcE&=&0,
    \label{mx2}
    \eea
where $k=\omega/c$ is the wavenumber, and $c=1/\sqrt{\varepsilon_0\mu_0}$ is the speed of light in vacuum. In terms of the scaled electric and magnetic fields, the time-averaged Poynting vector takes the form:
    \be
    \br\bS\kt=\frac{c}{2}\:\RE(\bcE\times\bcH^*),
    \ee
where ``Re'' stands for the real part of its argument.

The standard setup for the scattering of EM waves rests on the
assumption that for $r:=|\bbr| \to\infty$ the inhomogeneity of $\cM$
diminishes, i.e., $\hat\varepsilon(\bbr)\to 1$ and $\hat\mu(\bbr)\to
1$, at such a rate that (\ref{mx1}) and (\ref{mx2}) admit solutions
fulfilling the asymptotic boundary condition:
    \be
    \bcE(\bbr)= \cE_0 \left[ e^{i\bk_{\rm i}\cdot\bbr} \hat\bbe_{\rm i}+\frac{e^{ikr}}{r}\,f(\bk_{\rm s},\bk_{\rm i})\,\hat\bbe_{\rm s}\right]~~
    {\rm for}~~r\to\infty,
    \label{asym-BC}
    \ee
where $\cE_0$ is a constant, $\bk_{\rm i}$ and $\bk_{\rm s}:=k\bbr/r=k\hat\bbr$ are respectively the wave vectors for the incident and scattered waves, $\hat\bbe_{\rm i}$ and $\hat\bbe_{\rm s}$ are the unit vectors specifying the polarization of the incident and scattered waves, and $f(\bk_{\rm s},\bk_{\rm i})$ is the scattering amplitude whose modulus-square yields the differential cross section;
    \be
    \sigma_d(\bk_{\rm s},\bk_{\rm i})=\left|f(\bk_{\rm s},\bk_{\rm i})\right|^2.
    \label{diff-cross-sec}
    \ee
The first and second terms in the square bracket in (\ref{asym-BC}) respectively correspond to the incident and scattered waves;
    \begin{align}
    &\bcE_{\rm i}(\bbr):=\cE_0 e^{i\bk_{\rm i}\cdot\bbr} \hat\bbe_{\rm i},
    &&\bcE_{\rm s}(\bbr):= \frac{\cE_0\,e^{ikr}}{r}\,f(\bk_{\rm s},\bk_{\rm i})\,\hat\bbe_{\rm s}.
    \label{Ei-Es}
    \end{align}
Their wave and polarization vectors satisfy:
    \begin{align*}
    &|\bk_{\rm i}|=|\bk_{\rm s}|=k,
    &&\hat\bbe_{\rm i}\cdot\bk_{\rm i}=0,
    &&\hat\bbe_{\rm s}\cdot\bk_{\rm s}=k\,\hat\bbe_{\rm s}\cdot\hat\bbr=0.
    \end{align*}
Recall also that the differential cross section is defined in terms of the time-averaged Poynting vectors for the incident and scattered waves, $\br \bS_{\rm i}\kt$ and $\br \bS_{\rm s}\kt$, according to  \cite{TKD}:
    \be
    \sigma_d(\bk_{\rm s},\bk_{\rm i}):=\frac{r^2 |\br\bS_{\rm s}\kt|}{|\br\bS_{\rm i}\kt|}.
    \label{diff-cross-sec-def}
    \ee

Solving the scattering problem for $\cM$ means determining the scattering amplitude $f(\bk_{\rm s},\bk_{\rm i})$, which is generally a complex-valued function of $k$, the directions $\hat\bk_{\rm i}=\bk_{\rm i}/k$ and $\hat\bk_{\rm s}=\hat\bbr$ of the incident and scattered wavevectors, and their polarization $\hat\bbe_{\rm i}$ and $\hat\bbe_{\rm s}$.

\section{Transfer matrix for scattering of EM waves}
\label{S3}

Let us choose a coordinate system in which the source of the incident wave and the detectors
are placed on the planes $z=\pm\infty$. If the source is located at $z=-\infty$ (respectively $z=+\infty$) we use the qualification ``left-incident'' (respectively ``right-incident'') for the incident wave. In this case, $\hat \bk_{\rm i}\cdot\hat\bbe_z>0$ (respectively $\hat\bk_{\rm i}\cdot\hat\bbe_z<0$). In the following, we first consider the scattering problem for the left-incident waves, which we refer to as ``scattering from the left.''

We begin our analysis by introducing a useful notation: Given a vector or a vector-valued function, $\bv=v_x\bbe_x+v_y\bbe_y+v_z\bbe_z$, we use $\vec v$ to denote the projection of $\bv$ onto the $x$-$y$ plane, i.e.,
$\vec v:=v_x\bbe_x+v_y\bbe_y$, and identify it with the column vector $
    \left[\begin{array}{c}
    v_x \\
    v_y\end{array}\right]$.

\subsection{Four-component EM fields}

In view of (\ref{mx1}) and (\ref{mx2}), we can express $\cE_z$ and $\cH_z$ in the form
    \be
    \begin{aligned}
    &\cE_z=\frac{i}{k\hat\varepsilon}(\partial_x\cH_y-\partial_y\cH_x),\\
    &\cH_z=-\frac{i}{k\hat\mu}(\partial_x\cE_y-\partial_y\cE_x).
    \end{aligned}
    \label{mx3}
    \ee
Substituting these back into (\ref{mx1}) and (\ref{mx2}), we obtain a system of equations for the components of $\vec\cE:=\left[\begin{array}{c}
    \cE_x \\
    \cE_y\end{array}\right]$ and
    $\vec\cH:=\left[\begin{array}{c}
    \cH_x \\
    \cH_y\end{array}\right]$.
In terms of the four-component field \cite{berreman-1972,schubert-1996,pendry-1994},
    \begin{align}
    &\bPhi:=\left[\begin{array}{c}
    \cE_x \\
    \cE_y \\
    \cH_x \\
    \cH_y\end{array}\right]=\left[\begin{array}{c}
    \vec\cE \\
    \vec\cH \end{array}\right],
    \label{4-comp}
    \end{align}
this is equivalent to the Schr\"odinger equation:
    \be
    i\partial_z\bPhi(x,y,z)={\widehat{\bH}}\,\bPhi(x,y,z),
    \label{sch-eq-EM}
    \ee
where
    \be
    {\widehat{\bH}}:=\left[\begin{array}{cc}
    \bzero & {\widehat{\bL}}[\hat\varepsilon^{-1},\hat\mu]\\
    - {\widehat{\bL}}[\hat\mu^{-1},\hat\varepsilon] & \bzero\end{array}\right],
    \label{H=1}
    \ee
$\bzero$ is the $2\times 2$ null matrix, and for each pair of complex-valued functions $f(\bbr)$ and $g(\bbr)$,
    \be
    {\widehat{\bL}}[f,g]:=\frac{1}{k}\left[\begin{array}{cc}
    f\partial_x \partial_y+(\partial_x f)\partial_y  &
    -f\partial^2_x -(\partial_x f)\partial_x-k^2g\\
    f\partial^2_y +(\partial_y f)\partial_y+k^2g&
    -f\partial_y \partial_x
    -(\partial_y f)\, \partial_x\end{array}\right].
     \label{L=}
     \ee

For EM waves propagating in vacuum, where
$\hat\varepsilon=\hat\mu=1$, we have
${\widehat{\bH}}={\widehat{\bH}}_0$, where
    \be
    {\widehat{\bH}}_0:=\left[\begin{array}{cc}
    \bzero & {\widehat{\bL}}_0\\
    - {\widehat{\bL}}_0 & \bzero\end{array}\right],~~~~~~{\widehat{\bL}}_0:={\widehat{\bL}}[1,1].
    \label{H-zero}
    \ee
These together with (\ref{L=}) imply
    \be
    {\widehat{\bH}}_0^2=(\partial_x^2+\partial_y^2+k^2)\,\bI,
    \label{Hzero-2}
    \ee
where $\bI$ is the identity matrix.\footnote{Throughout this article we respectively use $\bzero$ and $\bI$ to denote the null and identity matrix of appropriate size.} Denoting the four-component field $\bPhi$ for waves propagating in vacuum by $\bPhi_0$, so that
    \be
    i\partial_z\bPhi_0={\widehat{\bH}}_0\bPhi_0,
    \label{sch-eq-free}
    \ee
applying ${\widehat{\bH}}_0$ to both sides of this equation from the left, and using (\ref{Hzero-2}), we find the Helmholtz equation: $ (\bnabla^2+k^2)\bPhi_0=0$. We can express the general plane-wave solutions of this equation in the form,
    \be
    \bPhi_0(\bbr)=\bPhi_0(\vec r,z)=\frac{1}{4\pi^2} \int_{\sD_k} d^2\vec p\: e^{i\vec p\cdot\vec r}
    \left[ \bA(\vec p) e^{i\varpi(\vec p)z}+ \bB(\vec p)e^{-i\varpi(\vec p)z}\right],
    \label{pw}
    \ee
where
    \begin{align*}
    &\vec p:=p_x\bbe_x+p_y\bbe_y,
    &&\vec r:=x\,\bbe_x+y\,\bbe_y,\\[6pt]
    &\sD_k:=\left\{~\vec p\in\R^2~\big|~|\vec p|< k~\right\},
    &&\varpi(\vec p):=\sqrt{k^2-|\vec p|^2},
    \end{align*}
and $\bA,\bB:\R^2\to\C^4$ are vector-valued coefficient functions that vanish outside $\sD_k$; they belong to the function space $\sF^4_k$, where
    \be
    \sF^d_k:=\left\{ \bF:\R^2\to\C^d~\big|~ \bF(\vec p)=0~{\rm for}~|\vec p|\geq k~\right\},~~~~~
    d=1,2,3,4.
    \label{func-space}
    \ee
In the following we identify elements $\bF$ of $\sF^d_k$ with $d$-component fields whose components belong to $\sF^1_k$. In particular, for $d=4$, we have
    \be
    \bF=\left[\begin{array}{c}
    F_1\\ F_2\\ F_3\\F_4\end{array}\right]~~{\rm and}~~
    F_1,F_2,F_3,F_4\in \sF^1_k.
    \label{4-comp-F}
    \ee

Next, we examine the Fourier transform of the plane-wave solution (\ref{pw}) with respect to $\vec r$, i.e.,
    \be
    \tilde\bPhi_0(\vec p,z)=\cF_{\vec p}\{\bPhi_0(\vec r,z)\}:=
    \int_{\R^2}d^2\vec r\: e^{-i\vec p\cdot\vec r}\bPhi_0(\vec r,z).
    \label{FT}
    \ee
It is clear from  (\ref{pw}) that
    \be
    \tilde\bPhi_0(\vec p,z)=\bA(\vec p)e^{i\varpi(\vec p)z}+\bB(\vec p)e^{-i\varpi(\vec p)z}.
    \label{t-pw}
    \ee
We can use this equation together with (\ref{sch-eq-free}) and (\ref{pw}) to show that
    \begin{align}
    &\tilde{\bH}_0(\vec p)\bA(\vec p)=-\varpi(\vec p)\bA(\vec p),
    &&\tilde{\bH}_0(\vec p)\bB(\vec p)=\varpi(\vec p)\bB(\vec p),
    \label{eg-va-eq}
    \end{align}
where
    \begin{align}
    &\tilde{\bH}_0(\vec p):=\left[\begin{array}{cc}
    \bzero & \tilde{\bL}_0(\vec p)\\
    - \tilde{\bL}_0(\vec p) & \bzero\end{array}\right],
    && \tilde{\bL}_0(\vec p):=\frac{1}{k}\left[\begin{array}{cc}
    -p_xp_y & p_x^2-k^2\\
    -p_y^2+k^2 & p_xp_y\end{array}\right].
    \label{t-H-zero}
    \end{align}
It is also easy to see that
    \be
    \tilde{\bH}_0(\vec p)^2=\varpi(\vec p)^2\bI.
    \label{Hzero-sq}
    \ee

According to Eq.~(\ref{eg-va-eq}) and the fact that $\varpi(\vec p)\neq 0$ for $\vec p\in\sD_k$, $\bA(\vec p)$ and $\bB(\vec p)$ are eigenvectors of $\tilde{\bH}_0(\vec p)$ with distinct eigenvalues. This implies that they are linearly independent. In particular, we can determine $\bA(\vec p)$ and $\bB(\vec p)$ from their sum, $\bC(\vec p):=\bA(\vec p)+\bB(\vec p)$, using the relations:
    \begin{align}
    &\bA(\vec p)=\bPi_1(\vec p)\bC(\vec p),
    &&\bB(\vec p)=\bPi_2(\vec p)\bC(\vec p),
    \label{id-1}
    \end{align}
where
    \be
    \bPi_j(\vec p):=\frac{1}{2}\left[\bI
    +\frac{(-1)^j}{\varpi(\vec p)}\tilde{\bH}_0(\vec p)\right],~~~~~~~~~~j=1,2.
    \label{proj}
    \ee
It is easy to see that for every $\bF\in\sF_k^4$, $\bPi_j(\vec p)\bF(\vec p)$ is an eigenvector of $\tilde{\bH}_0(\vec p)$ with eigenvalue $(-1)^j\varpi(\vec p)$, and
    \begin{align}
    &\bPi_1(\vec p)+\bPi_2(\vec p)=\bI,
    &&\bPi_i(\vec p)\bPi_j(\vec p)=\delta_{ij}\bPi_j(\vec p),
    \label{id-2}
    \end{align}
where $\delta_{ij}$ is the Kronecker delta symbol. Let us also note that in light of
(\ref{t-pw}), (\ref{eg-va-eq}), (\ref{id-1}), and (\ref{id-2}),
    \be
    \bC(\vec p)=e^{iz\tilde{\bH}_0(\vec p)}\tilde\bPhi_0(\vec p,z).
    \label{C=}
    \ee

Next, we identify $4\times 4$ matrices $\bK$ with the linear operators acting on the space $\C^4$ of $4\times 1$ complex matrices $\bX$ by matrix multiplication, i.e., $\bX\to\bK\bX$. Then (\ref{id-2}) identifies $\{\bPi_1(\vec p),\bPi_2(\vec p)\}$ with a complete orthogonal set of projection operators that project vectors onto the eigenspaces of $\tilde{\bH}_0(\vec p)$, i.e., they are eigenprojection operators of $\tilde{\bH}_0(\vec p)$. Because $\tilde{\bH}_0(\vec p)$ is manifestly non-Hermitian, the existence of a corresponding complete orthogonal set of eigenprojectors may seem unexpected. A simple explanation for this phenomenon is provided by the fact that $\tilde{\bH}_0(\vec p)$ is an $\bfeta_+$-pseudo-Hermitian operator, i.e., $\tilde{\bH}_0(\vec p)^\dagger=\bfeta_+\tilde{\bH}_0(\vec p)\bfeta_+^{-1}$, for a positive-definite matrix (metric operator) $\bfeta_+$, \cite{ppp1,review}. This follows from the fact that $\tilde{\bH}_0(\vec p)$ is a diagonalizable matrix with real eigenvalues and identifies it with a self-adjoint operator acting in the inner-product space $\C^4_{\bfeta_+}$ obtained by endowing $\C^4$ with the inner product $\br\bX,\bY\kt_{\bfeta_+}:=\bX^\dagger\bfeta_+\bY$, \cite{ppp1,review}. We can indeed use the prescription outlined in \cite{ppp1,review} to determine the general form of $\bfeta_+$. A particular example is
    \be
    \bfeta_+=\left[\begin{array}{cc}
    \bfTheta & \bzero\\
    \bzero & \bI\end{array}\right],
    \label{bfeta=}
    \ee
where 
    \[\bfTheta:=\frac{1}{k^2\varpi(\vec p)^2}\left[\begin{array}{cc}
        (k^2-p_y^2)^2+p_x^2p_y^2 & p_xp_y[k^2+\varpi(\vec p)^2]\\
    p_xp_y[k^2+\varpi(\vec p)^2] & (k^2-p_x^2)^2+p_x^2p_y^2\end{array}\right].\]

\subsection{Definition of transfer matrix}

Consider the scattering setup for EM waves and suppose that the
inhomogeneity of the medium decays so rapidly for $z\to\pm\infty$
that every solution of the Maxwell's equations~(\ref{mx1}) and
(\ref{mx2}) tend to a plane wave as $z\to\pm\infty$. This implies
that the four-component field $\bPhi$ has the following asymptotic
expression.
    \be
    \bPhi(\vec r,z)=\bPhi_\pm(\vec r,z)~~{\rm for}~~z\to\pm\infty,
    \label{asym-EM}
    \ee
where
    \be
    \bPhi_\pm(\vec r,z):=\frac{1}{4\pi^2} \int_{\sD_k} d^2\vec p\: e^{i\vec p\cdot\vec r}
    \left[ \bA_\pm(\vec p) e^{i\varpi(\vec p)z}+ \bB_\pm(\vec p)e^{-i\varpi(\vec p)z}\right],
    \label{asym-EM-2}
    \ee
and $\bA_\pm$ and $\bB_\pm$ belong to $\sF^4_k$. Because $\bPhi_\pm(\vec r,z)$ are plane-wave solutions of (\ref{sch-eq-free}), the sum of $\bA_\pm$ and $\bB_\pm$, i.e.,
    \be
    \bC_\pm:=\bA_\pm+\bB_\pm,
    \label{C-pm}
    \ee
satisfies
    \be
    \bC_\pm(\vec p)=e^{iz\tilde{\bH}_0(\vec p)}\tilde\bPhi_\pm(\vec p,z).
    \label{C=2}
    \ee

Let us now recall the definition of the transfer matrix in one dimension [Eq.~(\ref{M=1D})] and its higher dimensional
generalization that is given in Ref.~\cite{pra-2016}. These together with (\ref{asym-EM}) and  (\ref{asym-EM-2}) suggest to identity the EM transfer matrix as a linear operator that maps $\bA_-$ and $\bB_-$ to $\bA_+$ and $\bB_+$. We can introduce the operators $\widehat\bPi_j:\sF_k^4\to\sF_k^4$ by
    \begin{align}
    &(\widehat\bPi_j\bF)(\vec p):=\bPi_j(\vec p)\bF(\vec p)~~{\rm for~all}~~\bF\in\sF_k^4,
    \label{bPi-def}
    \end{align}
and express (\ref{id-1}) as
    \begin{align}
    &\bA_\pm=\widehat\bPi_1\bC_\pm,
    &&\bB_\pm=\widehat\bPi_2\bC_\pm.
    \label{AB-Pi}
    \end{align}
According to these equations, we can recover $\bA_\pm$ and $\bB_\pm$
from $\bC_\pm$. Motivated by this observation, we propose the
following definition for an EM transfer matrix.
    \begin{itemize}
    \item[]{\em Definition~1:} The {\em transfer matrix} for the electromagnetic waves scattered by an isotropic medium is the linear operator $\widehat{\bM}:\sF_k^4\to\sF_k^4$
    satisfying,
    \be
    \bC_+=\widehat{\bM}\bC_-,
    \label{M-EM}
    \ee
where $\bC_\pm:=\bA_\pm+\bB_\pm$, and $\bA_\pm$ and $\bB_\pm$ are the coefficient functions determining the asymptotic expression for the scattering solutions of (\ref{sch-eq-EM}) via (\ref{asym-EM}) and (\ref{asym-EM-2}).
    \end{itemize}
Because $\sF_k^4$ consists of four-component fields (\ref{4-comp-F}) with components belonging to $\sF_k^1$, $\widehat{\bM}$ is a $4\times 4$ matrix with operator entries acting in $\sF_k^1$; it is not a numerical matrix. We can identify it with an integral operator that has a complex $4\times 4$ matrix-valued kernel $\bM(\vec p,\vec q)$;
    \be
    \big(\widehat{\bM}\bF\big)(\vec p)=
    \int_{\sD_k}d^2\vec{q}\;\bM(\vec p,\vec q)\,\bF(\vec q).
    \nn
    \ee

The question of the existence and uniqueness of $\widehat{\bM}$ is
equivalent to whether the asymptotic expression for the
electromagnetic field at $z=-\infty$ determines the field and
consequently its asymptotic expression at $z=+\infty$ in a unique
manner. The latter is a physical condition that is clearly fulfilled
for situations where $\hat\varepsilon-1$ and $\hat\mu-1$ have
compact supports.

%\section{Evaluation of the scattering amplitude}
%\label{S4}

\subsection{Transfer matrix and the Reflection and transmission amplitudes}

Consider a left-incident plane wave with wavevector $\bk_{\rm i}$, polarization vector $\hat\bbe_{\rm i}$, and the scaled electric and magnetic fields:
    \begin{align}
    &\bcE_{\rm i}(\bbr)=\cE_0\, e^{i\bk_{\rm i}\cdot\bbr}\,\hat\bbe_{\rm i},
    &&\bcH_{\rm i}(\bbr)=\cE_0\, e^{i\bk_{\rm i}\cdot\bbr}\,\hat\bk_{\rm i}\times\hat\bbe_{\rm i}.
    \label{left-EH}
    \end{align}
Because $\bk_{\rm i}$ has a positive component along the $z$-axis, we can express it in the form
    \be
    \bk_{\rm i}=\vec k_{\rm i}+\varpi(\vec k_{\rm i})\,\hat\bbe_z,
    \ee
where $\vec k_{\rm i}$ is the projection of $\bk_{\rm i}$ onto the $x$-$y$ plane, and
    \be
    \varpi(\vec k_{\rm i}):=\sqrt{k^2-|\vec k_{\rm i}|^2}.\nn
    \ee

In view of (\ref{left-EH}), the four-component field (\ref{4-comp}) for the above left-incident wave is given by
    \begin{align}
    &\bPhi_{\rm i}(\bbr)=\cE_0\, e^{i\bk_{\rm i}\cdot\bbr}\,\bup_{\rm i},
    \label{4-comp-i}
    \end{align}
where
    \begin{align}
    &\bup_{\rm i}:=\left[\begin{array}{c}
    \hat\bbe_x\cdot\hat\bbe_{\rm i}\\
    \hat\bbe_y\cdot\hat\bbe_{\rm i}\\
    \hat\bbe_x\cdot(\hat\bk_{\rm i}\times\hat\bbe_{\rm i})\\
    \hat\bbe_y\cdot(\hat\bk_{\rm i}\times\hat\bbe_{\rm i})\end{array}\right]
    =\left[\begin{array}{c}
    \hat\bbe_x\cdot\hat\bbe_{\rm i}\\
    \hat\bbe_y\cdot\hat\bbe_{\rm i}\\
    (\hat\bbe_x\times\hat\bk_{\rm i})\cdot\hat\bbe_{\rm i}\\
    (\hat\bbe_y\times\hat\bk_{\rm i})\cdot\hat\bbe_{\rm i}\end{array}\right].
    \label{Upsilon-i=}
    \end{align}
Equation~(\ref{4-comp-i}) together with the fact that in the scattering process for a
left-incident wave there is no wave emitted from a source located on
the plane $z=+\infty$ show that the scattering solution
(\ref{asym-BC}) corresponds to the following choice for the
coefficient functions $\bA_-(\vec p)$ and $\bB_+(\vec p)$ entering
(\ref{asym-EM-2}):
    \begin{align}
    &\bA_-(\vec p)=4\pi^2\delta(\vec p-\vec k_{\rm i})\bup_{\rm i},
    &&\bB_+(\vec p)=\bzero.
    \label{ini-condi}
    \end{align}
Furthermore, according to (\ref{id-2}) and (\ref{AB-Pi}),
    \begin{align}
    &\bPi_1(\vec p)\bA_\pm(\vec p)=\bA_\pm(\vec p),
    &&\bPi_2(\vec p)\bB_\pm(\vec p)=\bB_\pm(\vec p),
    &&\bPi_1(\vec p)\bB_\pm(\vec p)=\bPi_2(\vec p)\bA_\pm(\vec p)=\bzero.
    \label{id-5}
    \end{align}
The first of these relations together with (\ref{ini-condi}) imply
    \begin{align}
    &\bPi_1(\vec k_{\rm i})\bup_{\rm i}=\bup_{\rm i},
    &&\bPi_2(\vec k_{\rm i})\bup_{\rm i}=\bzero.
    \label{id-6}
    \end{align}

Next, we introduce:
    \begin{align}
    &\bT_-:=\bB_--\bB_+,
    &&\bT_+:=\bA_+-\bA_-.
    \label{Ts}
    \end{align}
For the scattering of a left-incident plane wave, where
(\ref{ini-condi}) holds, $\bT_\pm$ yield the left reflection
and transmission amplitudes:
    \begin{align}
    &\bT^l_-(\vec p):=\bB_-(\vec p),
    &&\bT^l_+(\vec p):=\bA_+(\vec p)-4\pi^2\delta(\vec p-\vec k_{\rm i})\bup_{\rm i}.
    \label{left-Ts}
    \end{align}
To see the reason for this terminology, consider the asymptotic form
of the four-component field for the scattered wave which we label by
$\bPhi_{\rm s}(\bbr)$. According to (\ref{asym-BC}) and (\ref{Ei-Es}), the scaled
electric field for the scattered wave has the form
$\bcE(\bbr)-\bcE_{\rm i}(\bbr)$ for $r\to\infty$. This shows that
    \be
    \bPhi_{\rm s}(\bbr)=\bPhi(\bbr)-\bPhi_{\rm i}(\bbr),
    \label{P=P-P}
    \ee
where $\bPhi(\bbr)$ is the four-component field given by
(\ref{asym-EM}), (\ref{asym-EM-2}), and (\ref{ini-condi}).
Evaluating the Fourier transform of the both sides of (\ref{P=P-P})
with respect to $\vec r$, taking the limit $z\to\pm\infty$, and
using (\ref{left-Ts}), we find
    \be
    \tilde\bPhi_{\rm s}(\vec p,z)=\left\{\begin{array}{ccc}
    \bT^l_-(\vec p) e^{-i\varpi(\vec p)z} & {\rm for} & z\to-\infty,\\[6pt]
    \bT^l_+(\vec p) e^{i\varpi(\vec p)z} & {\rm for} & z\to+\infty.
    \end{array}\right.
    \label{tPsi-s=}
    \ee
Let us also note the following simple consequences of (\ref{bPi-def}), (\ref{id-5}), (\ref{id-6}), and (\ref{left-Ts}).
    \begin{align}
    &\widehat\bPi_1\bT^l_+=\bT^l_+,
    &&\widehat\bPi_2\bT^l_-=\bT^l_-,
    &&\widehat\bPi_1\bT^l_-=\widehat\bPi_2\bT^l_+=\bzero.
    \label{id-21}
    \end{align}

With the help of (\ref{ini-condi}) and (\ref{left-Ts}), we can express the coefficient functions $\bC_\pm$ of (\ref{C-pm}) in the form:
    \be
    \bC_\pm=\bT^l_\pm+4\pi^2\bup_{\rm i}\delta_{\vec k_{\rm i}},
    \label{Cpm=z1}
    \ee
where $\delta_{\vec k_{\rm i}}$ stands for the Dirac delta function centered at $\vec k_{\rm i}$, i.e.,
    \be
    \delta_{\vec k_{\rm i}}(\vec p):=\delta(\vec p-\vec k_{\rm i}).
    \ee
Substituting (\ref{Cpm=z1}) in (\ref{M-EM}), we find
    \be
    \bT^l_+=\widehat{\bM}\bT^l_-+4\pi^2(\widehat{\bM}-\widehat\bI)\bup_{\rm i}\delta_{k_{\rm i}}.
    \label{Tp=}
    \ee
Here and in what follows $\widehat\bI$ stands for the identity operator acting in $\sF^d_k$, i.e., for all $\bF\in\sF_k^d$, $\widehat\bI\bF:=\bF=\bI\bF$. If we apply the projection operator $\widehat\bPi_2$ of (\ref{bPi-def}) to both sides of (\ref{Tp=}) and make use of (\ref{id-6}) and (\ref{id-21}), we obtain
    \be
    \widehat\bPi_2\widehat{\bM}\bT^l_-=-4\pi^2\widehat\bPi_2\widehat{\bM}\bup_{\rm i} \delta_{\vec k_{\rm i}}.
    \label{Tm=}
    \ee
Furthermore, using (\ref{id-21}) we can respectively establish the following consequences of (\ref{Tp=}) and (\ref{Tm=}).
    \bea
    \bT^l_+&=&-\widehat\bPi_1\left(\widehat\bI-\widehat{\bM}\right)\left(\bT^l_-+4\pi^2
    \bup_{\rm i} \delta_{\vec k_{\rm i}}\right),
    \label{d-eq02}\\
    \bT^l_-&=&\widehat\bPi_2\left(\widehat\bI-\widehat{\bM}\right)\left(\bT^l_-+4\pi^2
    \bup_{\rm i} \delta_{\vec k_{\rm i}}\right).
    \label{d-eq01}
    \eea

Equations~(\ref{Tp=}) and (\ref{Tm=}) turn out to be equivalent to a pair of equations that have the same structure as those satisfied by the entries of the transfer matrix of potential scattering in two and three dimensions \cite{pra-2016}.
To see this we introduce the operators $\widehat{\bM}_{ij}:\sF_k^4\to\sF_k^4$ according to
    \be
    \widehat{\bM}_{ij}:=\widehat\bPi_i\widehat{\bM}\,\widehat\bPi_j,
    \quad\quad\quad i,j=1,2,
    \ee
and employ (\ref{id-21}) to express (\ref{Tm=}) as
    \be
    \widehat{\bM}_{22}\bT^l_-=-4\pi^2\widehat{\bM}_{21}\bup_{\rm i} \delta_{\vec k_{\rm i}}.
    \label{Tm=2}
    \ee
Similarly by applying $\widehat\bPi_1$ to both sides of (\ref{Tp=}), we have
    \be
    \bT^l_+=\widehat{\bM}_{12}\bT^l_-+4\pi^2(\widehat{\bM}_{11}-\widehat\bI)\bup_{\rm i} \delta_{\vec k_{\rm i}}.
    \label{Tp=2}
    \ee
Equation~(\ref{Tm=2}) is a system of linear non-homogeneous integral equations for the left reflection amplitude $\bT^l_-(\vec p)$. Solving this system and inserting the result in (\ref{Tp=}) we can determine the left transmission amplitude $\bT^l_+(\vec p)$. Expressing the solution of  (\ref{Tp=}) as the application of the inverse of the operator $\widehat{\bM}_{22}$ on the right-hand side of this equation and using the result in (\ref{Tp=2}), we find
    \bea
    \bT^l_-&=&-4\pi^2\widehat{\bM}_{22}^{-1}\widehat{\bM}_{21}\bup_{\rm i} \delta_{k_{\rm i}},
    \label{Tm=3}\\
    \bT^l_+&=&4\pi^2\left(\widehat{\bM}_{11}-\widehat\bI-\widehat{\bM}_{12}
    \widehat{\bM}_{22}^{-1}\widehat{\bM}_{21}\right)\!\bup_{\rm i} \delta_{k_{\rm i}}.
    \label{Tp=3}
    \eea
 It is absolutely remarkable that dropping $\bup_{\rm i}$ in (\ref{Tm=3}) and (\ref{Tp=3}) we recover equations~(20) of Ref.~\cite{pra-2016} which are derived for the entries of the transfer matrix of potential scattering in two and three dimensions.

\subsection{Connection to scattering amplitude and cross section}

In the preceding subsection, we show that the reflection and transmission amplitudes $\bT^l_\pm(\vec p)$ satisfy a set of linear equations involving the EM transfer matrix $\widehat{\bM}$. Here we derive explicit expressions for the scattering amplitude and differential cross section in terms of $\bT^l_\pm(\vec p)$. This in turn establishes the physical significance of our transfer matrix as a linear (integral) operator containing the complete information about the scattering features of the scattering medium.

First, we consider the four-component field for the scattered wave,
    \be
    \bPhi_{\rm s}(\bbr)=\left[\begin{array}{c}
    \vec\cE_{\rm s}(\bbr)\\
    \vec\cH_{\rm s}(\bbr)\end{array}\right],
    \label{bPhi-s}
    \ee
and recall that in view of (\ref{mx1}), (\ref{Ei-Es}), and the fact
that $\hat\mu=1$ for $r\to\infty$,
    \begin{align}
    &\bcE_{\rm s}(\bbr)=\frac{\cE_0\,e^{ikr}}{r}\,f(\bk_{\rm s},\bk_{\rm i})\,\hat\bbe_{\rm s} 
    ~~~~{\rm for}~~~~r\to\infty,
    \label{Es=4}\\
    &\bcH_{\rm s}(\bbr)=\frac{\cE_0\,e^{ikr}}{r}\,f(\bk_{\rm s},\bk_{\rm i})\,\hat r\times\hat\bbe_{\rm s}
    ~~~~{\rm for}~~~~r\to\infty.
    \label{Hs=4}
    \end{align}
We can also obtain the following asymptotic expression for $\bPhi_{\rm s}(\bbr)$ by evaluating the inverse Fourier transform of both sides of (\ref{tPsi-s=}) with respect to $\vec p$.
    \be
    \bPhi_{\rm s}(\vec r,z)=\frac{1}{4\pi^2}\int_{\sD_k}d^2\vec p
    \;\bT^l_\pm(\vec p) e^{\pm i\varpi(\vec p)z} e^{i\vec p\cdot\vec r}
    ~~~{\rm for}~~~z\to\pm\infty.
    \label{Psi-s=}
    \ee
Using the asymptotic expression for $e^{\pm i\varpi(\vec p)z} e^{i\vec p\cdot\vec r}$ that is derived in Appendix~F of Ref.~\cite{pra-2016}, we can write (\ref{Psi-s=}) in the form:
    \be
    \bPhi_{\rm s}(\bbr)=-\frac{i\cE_e e^{ikr}}{2\pi r}\,\varpi(\vec k_{\rm s})
    \bT^l_\pm(\vec k_{\rm s})
    ~~~{\rm for}~~~z\to\pm\infty.
    \label{bPhi-s2}
    \ee

Substituting (\ref{Es=4}) and (\ref{Hs=4}) in (\ref{bPhi-s}) and comparing the result with (\ref{bPhi-s2}), we find
    \be
    f(\bk_{\rm i},\bk_{\rm s})\bup_{\rm s}=-\frac{i\varpi(\vec k_{\rm s})}{2\pi}\bT^l_\pm(\vec k_{\rm s})~~~{\rm for}~~~
    \pm\hat\bbe_z\cdot\hat\bk_{\rm s}>0,
    \label{f=1}
    \ee
where
    \bea
    &&\bup_{\rm s}:=\left[\begin{array}{c}
    \hat\bbe_x\cdot\hat\bbe_{\rm s}\\
    \hat\bbe_y\cdot\hat\bbe_{\rm s}\\
    \hat\bbe_x\cdot(\hat\bbr\times\hat\bbe_{\rm s})\\
    \hat\bbe_y\cdot(\hat\bbr\times\hat\bbe_{\rm s})\end{array}\right]
    =\left[\begin{array}{c}
    \hat\bbe_x\cdot\hat\bbe_{\rm s}\\
    \hat\bbe_y\cdot\hat\bbe_{\rm s}\\
    (\hat\bbe_x\times\hat\bbr)\cdot\hat\bbe_{\rm s}\\
    (\hat\bbe_y\times\hat\bbr)\cdot\hat\bbe_{\rm s}
    \end{array}\right]=
    \left[\begin{array}{c}
    \hat\bbe_x\cdot\hat\bbe_{\rm s}\\
    \hat\bbe_y\cdot\hat\bbe_{\rm s}\\
    (-\cos\vartheta\: \hat\bbe_y+
    \sin\vartheta\sin\varphi\:\hat\bbe_z)\cdot\hat\bbe_{\rm s}\\
    (\cos\vartheta\: \hat\bbe_x
    -\sin\vartheta\cos\varphi\:\hat\bbe_z)\cdot\hat\bbe_{\rm s}
    \end{array}\right],
    \label{Upsilon-s=}
    \eea
and $\vartheta$ and $\varphi$ are respectively the polar and azimuthal angles in the spherical coordinates, so that
    \[\hat\bk_{\rm s}=\hat\bbr=\sin\vartheta\cos\varphi\,\hat\bbe_x+
    \sin\vartheta\sin\varphi\,\hat\bbe_y+
    \cos\vartheta\,\hat\bbe_z.\]
We can use~(\ref{Upsilon-s=}) to determine $\hat\bbe_s$ in  terms of
$\bup_{\rm s}$. To see this, we introduce
    \begin{align}
    &\bEpsilon_x:=\left[\begin{array}{c}
    1 \\ 0 \\ 0 \\ 0 \end{array}\right],
    &&\bEpsilon_y:=\left[\begin{array}{c}
    0 \\ 1 \\ 0 \\ 0\end{array}\right],
    &&\bEpsilon_z:=\left[\begin{array}{c}
    0 \\ 0 \\ \sin\vartheta\sin\varphi\\
    -\sin\vartheta\cos\varphi\end{array}\right],
    \nn
    \end{align}
and use $\hat\bbr\cdot\hat\bbe_{\rm s}=0$ to check that
$\bEpsilon_j^\dagger\bup_{\rm s}=\hat\bbe_j\cdot\hat\bbe_{\rm s}$
for $j=x,y,z$. The latter equation implies
    \be
    \hat\bbe_{\rm s}=\bXi^\dagger\bup_{\rm s},
    \label{d-bXi}
    \ee
where
    \be
    \bXi^\dagger:=\hat\bbe_x\bEpsilon_x^\dagger+\hat\bbe_y
    \bEpsilon_y^\dagger+\hat\bbe_z\bEpsilon_z^\dagger.
    \label{d-bXi=}
    \ee
Applying $\bXi^\dagger$ to both sides of (\ref{f=1}) from the left
and using (\ref{d-bXi}) yield
    \be
    f(\bk_{\rm i},\bk_{\rm s})\hat\bbe_{\rm s}=
    -\frac{i\varpi(\vec k_{\rm s})}{2\pi}
    \bXi^\dagger\bT^l_\pm(\vec k_{\rm s})~~~{\rm for}~~~
    \pm\hat\bbe_z\cdot\hat\bk_{\rm s}>0.
    \label{fe=1}
    \ee
Dividing both sides of this equation by the norm of its right-hand
side and noting that $\hat\bbe_{\rm s}$ is a unit vector, we can
determine it up to a phase factor. Substituting the result in
(\ref{fe=1}) yields $f(\bk_{\rm i},\bk_{\rm s})$. The undetermined
phase factor is physically irrelevant, because it does not enter the
expression (\ref{Es=4}) for the scattered electric field. The latter
is uniquely determined by the right-hand side of (\ref{fe=1}).

Another consequence of (\ref{Upsilon-s=}) is the identity:
$\bup_{\rm s}^\dagger\bup_{\rm s}=1+\cos^2\vartheta$. This  together
with (\ref{f=1}) imply
    \begin{align}
    &f(\bk_{\rm i},\bk_{\rm s})=
    -\frac{i\varpi(\vec k_{\rm s})}{2\pi\sqrt{1+\cos^2\vartheta}}\hat\bup_{\rm s}^\dagger\bT^l_\pm(\vec k_{\rm s})
    ~~~{\rm for}~~~\pm\cos\vartheta>0,
    \label{f=2}
    \end{align}
where $\hat\bup_{\rm s}:=(1+\cos^2\vartheta)^{-1/2}\;\bup_{\rm s}$. According to (\ref{diff-cross-sec}) and (\ref{f=2}), the differential cross section is given by
    \be
    \sigma_d(\bk_{\rm i},\bk_{\rm s})=\frac{\varpi(\vec k_{\rm s})^2 \bT^l_\pm(\vec k_{\rm s})^\dagger
    \bT^l_\pm(\vec k_{\rm s})}{4\pi^2 (1+\cos^2\vartheta)}~~~{\rm for}~~~
    \pm\cos\vartheta>0,
    \label{cross-sec=1}
    \ee
where we have made use of the fact that $\bT^l_\pm(\vec k_{\rm s})$
is a scalar multiple of $\hat\bup_{\rm s}$, which is a unit
four-component vector. In Appendix~A we offer an alternative
derivation of (\ref{cross-sec=1}).

The above analysis reduces the solution of the scattering problem
for left-incident EM waves scattered by an isotropic medium to the
determination of the transfer matrix $\widehat{\bM}$ and the solution
of the integral equation for $\bT^l_-$, namely (\ref{Tm=2}).
This together with  (\ref{Tp=2}), (\ref{f=2}), and
(\ref{cross-sec=1}) yield the scattering amplitude and differential
cross section.

Now, consider a right-incident wave. Then the incident wavevector
$\bk_{\rm i}$ has a negative $z$-component, so that $\bk_{\rm
i}=\vec k_i-\varpi(\vec k)\hat\bbe_z$, and the coefficient functions
$\bA_-$ and $\bB_+$ appearing in the asymptotic expression for the
four-component field (\ref{asym-EM-2}) satisfy
    \begin{align}
    &\bA_-(\vec p)=\bzero,
    &&\bB_+(\vec p)=4\pi^2\delta(\vec p-\vec k_{\rm i})\bup_{\rm i}.
    \label{ini-condi-r}
    \end{align}
Substituting these in (\ref{Ts}), we find $\bT_\pm(\vec p)=\bT^r_\pm(\vec p)$ where
    \begin{align}
    &\bT^r_-(\vec p):=\bB_-(\vec p)-4\pi^2\delta(\vec p-\vec k_{\rm i})\bup_{\rm i},
    &&\bT^r_+(\vec p):=\bA_+(\vec p).
    \label{right-Ts}
    \end{align}
We can respectively interpret these as the transmission and reflection amplitudes for the right-incident wave. In view of (\ref{id-1}) and (\ref{ini-condi-r}), we have $\bPi_1(\vec k_{\rm i})\bup_{\rm i}=\bzero$ and $\bPi_2(\vec k_{\rm i})\bup_{\rm i}=\bup_{\rm i}$. Making use of these relations, we can repeat our derivation of the relationship between $\widehat{\bM}$ and $\bT^l_\pm$ to obtain the analogs of (\ref{Tm=2}) and (\ref{Tp=2}) for right-incident waves. This results in
    \begin{align}
    &\widehat{\bM}_{22}\bT^r_-=4\pi^2\left(\widehat\bI-\widehat{\bM}_{22}\right)\!\bup_{\rm i}
    \delta_{\vec k_{\rm i}},
    \label{Tm=2r}\\
    &\bT^r_+=\widehat{\bM}_{12} (\bT^r_-+4\pi^2\bup_{\rm i}\delta_{\vec k_{\rm i}}).
    \label{Tp=2r}
    \end{align}
Writing the solution of (\ref{Tm=2r}) in terms of the inverse of the operator $\widehat{\bM}_{22}$ and substituting the result in (\ref{Tp=2r}), we find the following analogs of (\ref{Tm=3}) and (\ref{Tp=3}).
    \begin{align}
    &\bT^r_-=4\pi^2\left(\widehat{\bM}_{22}^{-1}-\widehat\bI\right)\!\bup_{\rm i}\delta_{\vec k_{\rm i}},
    &&\bT^r_+=4\pi^2\widehat{\bM}_{12}\widehat{\bM}_{22}^{-1}\bup_{\rm i}\delta_{\vec k_{\rm i}}.
    \label{Tpm-3r}
    \end{align}
These equations have remarkably the same form as their analogs for the scattering of scalar waves \cite{prsa-2016}.

For a right-incident wave, the four-component field corresponding to
the scattered wave is given by (\ref{bPhi-s2}) with $\bT^l_\pm$
changed to $\bT^r_\pm$. Because (\ref{f=1})-- (\ref{cross-sec=1})
follow from (\ref{bPhi-s2}), we can use them for right-incident
waves provided that we change $\bT^l_\pm$ to $\bT^r_\pm$. In
particular, making this change in (\ref{f=2}) and
(\ref{cross-sec=1}) we find the scattering amplitude and
differential cross section for a right-incident wave.

\section{S-matrix description of the EM transfer matrix and its composition property}
\label{S5}

In this section, we show that the transfer matrix $\widehat{\bM}$ for
EM scattering can be expressed as the S-matrix of an effective
quantum system, i.e., it satisfies (\ref{M=exp}) for some Hamiltonian operator $\sH$. This in particular allows us to establish the
composition property for the EM transfer matrices that generalizes
(\ref{composition}).

First, we examine the Fourier transform of the four-component field $\bPhi(\vec r,z)$ with
respect to $\vec r $, i.e.,
    \[\tilde\bPhi(\vec\fK,z)=\cF_{\vec\fK}\{\bPhi(\vec r,z)\}:=
    \int_{\R^2}d^2\vec r\:e^{-i\vec r\cdot\vec\fK}\:\bPhi(\vec r,z).\]
We can use (\ref{sch-eq-EM}) to show that it satsifies:
    \be
    i\partial_z\tilde\bPhi(\vec \fK,z)=\widehat{\tilde{\bH}}(z)\tilde\bPhi(\vec \fK,z),
    \label{t-sch-eq-EM}
    \ee
where
    \bea
   \widehat{\tilde{\bH}}(z)&:=&\cF_{\vec\fK}\,{\widehat{\bH}}\,\cF^{-1}_{\vec r}=
    \left[\begin{array}{cc}
    \bzero &\widehat{ \tilde{\bL}}[\hat\varepsilon^{-1},\hat\mu]\\
    -\widehat{ \tilde{\bL}}[\hat\mu^{-1},\hat\varepsilon] & \bzero\end{array}\right]\,,
    \label{tH=1}\\
   \widehat{ \tilde{\bL}}[f,g]&:=&\cF_{\vec\fK}\,{\widehat{\bL}}[f(\vec r,z),g(\vec r,z)]\,\cF^{-1}_{\vec r}
    ={\widehat{\bL}}[f(\vec r,z),g(\vec r,z)]
    \Big|_{\vec r\to i\vec\nabla_{\fK},\vec\nabla_{r}\to i\vec\fK}\;,
    \label{tL}
    \eea
$\cF^{-1}_{\vec r}$ denotes inverse Fourier transformation with
respect  to $\vec\fK$, i.e., 
	\[\cF^{-1}_{\vec r}\{\psi(\vec\fK)\}:=
	\frac{1}{4\pi^2}\!\int_{\R^2}\! d^2\vec\fK\,e^{i\vec\fK\cdot\vec r}\psi(\vec\fK),\]
$\vec\nabla_{\fK}:=\bbe_x\partial_{\fK_x}+\bbe_y\partial_{\fK_y}$, $\vec\nabla_{r}:=\bbe_x\partial_{x}+\bbe_y\partial_{y}$, and the ${\widehat{\bL}}[f(\vec r,z),g(\vec r,z)]$
appearing in (\ref{tL}) is the normal-ordered operator given by (\ref{L=}). The entries of this operator are second-order differential operators with the following general structure:
    \[\widehat{\cL}(\vec r,\vec\nabla_r):=\sum_{i,j=x,y} f_{ij}(\vec r)\partial_i\partial_j+
    \sum_{i=x,y} g_{i}(\vec r)\partial_i+h(\vec r).\]
Therefore the computation of ${\widehat{\tilde{\bH}}}(z)$ involves
evaluating  operators of the form: $\cF_{\vec\fK}\,\widehat{\cL}(\vec
r,\vec\nabla_r)\,\cF^{-1}_{\vec r}$. Applying this operator on a
test function $\bxi$ and using the definition of $\cF_{\vec\fK}$ and
$\cF^{-1}_{\vec r}$, we have
    \be
    \big[\cF_{\vec\fK}\,\widehat{\cL}(\vec r,\vec\nabla_r)\,\cF^{-1}_{\vec r}\bxi\big](\vec\fK)=
    \big[ \cF_{\vec\fK}\,\widehat{\cL}(\vec r,\vec\nabla_r)\,\cF^{-1}_{\vec r}\big]\{\bxi(\vec q)\}=
    \frac{1}{4\pi^2}\int_{\R^2}d^2\vec q\: \tilde\fS_\cL(\vec\fK-\vec q,i\vec q)\,\bxi(\vec q),
    \label{integral-op1}
    \ee
where
    \be
    \tilde\fS_\cL(\vec\fK,i\vec q):=\int_{\R^2}d^2\vec r\:e^{-i\vec\fK\cdot\vec r}
    \fS_\cL(\vec r,i\vec q)
    \label{t-symbol}
    \ee
is the Fourier transform with respect to $\vec r$ of
    \be
    \fS_\cL(\vec r,i\vec q):=-\sum_{i,j=x,y} f_{ij}(\vec r)q_iq_j+
    i\sum_{i=x,y} g_{i}(\vec r)q_i+h(\vec r).
    \label{symbol}
    \ee
The latter is known as the symbol of $\widehat{\cL}(\vec r,\vec\nabla_r)$. 

For each value of $z$, the four-component field $\tilde\bPhi(\vec k,z)$ defines a function $\tilde\bPhi(\cdot,z):\R^2\to\C^4$, which for brevity we denote by
$\tilde\bPhi(z)$. This allows us to express (\ref{t-sch-eq-EM}) as
    \be
     i\partial_z\tilde\bPhi(z)=\widehat{\tilde{\bH}}(z)\tilde\bPhi(z).
        \label{t-sch-eq-EM2}
        \ee
This is a time-dependent Schr\"odinger equation with $z$ playing the
role of time. Let $z_0$ be an initial value of $z$, and
$\widehat{\tilde\bU}(z,z_0)$ denote the evolution operator associated with the
Hamiltonian $\widehat{\tilde{\bH}}(z)$. By definition, it satisfies
    \begin{align}
    &i\partial_z\widehat{\tilde\bU}(z,z_0)=\widehat{\tilde{\bH}}(z)\widehat{\tilde\bU}(z,z_0),
    \quad\quad\quad\widehat{\tilde\bU}(z_0,z_0)=\widehat\bI,
    \label{tU=}\\
    &\tilde\bPhi(z)=\widehat{\tilde\bU}(z,z_0)\tilde\bPhi(z_0).
    \label{tU=2}
    \end{align}
A simple consequence of
(\ref{asym-EM}) and (\ref{tU=2}) is
    \be
    \lim_{z\to\infty}\bPhi_+(z)=
    \lim_{z\to\infty}\lim_{z_0\to-\infty}\widehat{\tilde\bU}(z,z_0)\bPhi_-(z_0).
    \ee
We can use this equation together with (\ref{C=2}) to show that
    \bea
    \bC_+&=&\lim_{z\to\infty}
    e^{iz\widehat{\tilde{\bH}}_0}\tilde\bPhi_+(z)
    =\left[\lim_{z\to\infty}\lim_{z_0\to-\infty}
    e^{iz\widehat{\tilde{\bH}}_0}\widehat{\tilde\bU}(z,z_0)e^{-iz_0\widehat{\tilde{\bH}}_0}\right]
    \bC_-\nn\\
    &=&\widehat{\sU}(\infty,-\infty)\bC_-,
    \label{C=UC}
    \eea
where $\widehat{\tilde{\bH}}_0:\sF_k^4\to\sF_k^4$ is the linear
operator defined by
   \be
    (\widehat{\tilde\bH}_0\bF)(\vec p):=\tilde\bH_0(\vec p)\bF(\vec p)~~{\rm for}~~\bF\in\sF_k^4,
    \label{H-zero-op-def}
    \ee
$\tilde\bH_0(\vec p)$ is the $4\times 4$ matrix given by (\ref{t-H-zero}),
    \[\widehat{\sU}(z,z_0):=e^{iz\widehat{\tilde{\bH}}_0}
    \widehat{\tilde\bU}(z,z_0)e^{-iz_0\widehat{\tilde{\bH}}_0}=
    \sT\exp\left[-i\int_{z_0}^z dz'~\widehat{\sH}(z')\right]\]
is the evolution operator for the interaction-picture Hamiltonian:
    \be
    \widehat{\sH}(z):=e^{iz\widehat{\tilde{\bH}}_0}\left[\widehat{\tilde{\bH}}(z)-\widehat{\tilde{\bH}}_0\right]e^{-iz\widehat{\tilde{\bH}}_0},
    \label{int-pic-H}
    \ee
and $\sT$ stands for  the ``time-ordering'' operation with $z$
playing the role of ``time.'' The operator $\widehat{\sU}(\infty,-\infty)$ is
known as the scattering operator (S-matrix) \cite{weinberg} for the
effective quantum system $\cS$ determined by the Hamiltonian
operator $\widehat{\tilde{\bH}}(z)$. Because $\bC_\pm$ belong to the function
space $\sF^4_k$, we should think of $\widehat{\tilde{\bH}}(z)$ and $\widehat{\sH}(z)$ as
operators acting in $\sF^4_k$. They are indeed integral operators
whose integral kernels we derive in Appendix~B.

Comparing (\ref{M-EM}) and (\ref{C=UC}), we  are led to the
following remarkable result.
    \begin{itemize}
    \item[]{\em Theorem~1:} The electromagnetic transfer matrix coincides with the
    S-matrix of $\cS$, i.e.,
        \be
        \begin{aligned}
        \widehat{\bM}&=\widehat{\sU}(\infty,-\infty)=\sT\exp\left[-i\int_{-\infty}^\infty dz~\widehat{\sH}(z)\right]\\
%        &=\bI-i\int_{-\infty}^\infty \!\!dz_1~\widehat{\sH}(z_1)-
%        \int_{-\infty}^\infty \!\!dz_2\int_{-\infty}^{z_2} \!\!dz_1\,
%        \widehat{\sH}(z_2)\widehat{\sH}(z_1)+\cdots\\
        &=\widehat\bI+\sum_{\ell=1}^\infty (-i)^\ell
        \int_{-\infty}^\infty \!\!dz_\ell\int_{-\infty}^{z_\ell} \!\!dz_{\ell-1}
        \cdots\int_{-\infty}^{z_2} \!\!dz_1\,
        \widehat{\sH}(z_\ell)\widehat{\sH}(z_{\ell-1})\cdots\widehat{\sH}(z_1).
        \end{aligned}
        \label{M=exp3}
        \ee
    \end{itemize}
A straightforward consequence of this theorem is the composition property of the EM transfer matrix.

Consider the scattering of EM waves by an isotropic medium $\cM$.
Let us divide the space by $n-1$ planes that are orthogonal to the
$z$-axis and intersect it at $z=a_1$, $z=a_2,\cdots, z=a_{n-1}$.
This yields a partition of the space into $n$ regions that we
identify with their intersection with the $z$-axis, namely
$I_1:=(a_0,a_1]$, $I_2:=(a_1,a_2]$, $\cdots$,
$I_{n-1}:=(a_{n-2},a_{n-1}]$, and $I_{n}:=(a_{n-1},a_n)$, where
$a_0:=-\infty$ and $a_{n}:=\infty$. Clearly, along the $z$-axis,
$I_i$ is to the left of $I_j$ if and only if $i<j$.
%We denote this by $I_i< I_j$.
Let $\cM_\ell$ be a medium whose electromagnetic
properties are identical to those of $\cM$ in the region $I_\ell$
and coincides with vacuum outside $I_\ell$. We refer to $I_\ell$ as
the ``support'' of $\cM_\ell$. The interaction-picture Hamiltonian $\widehat{\sH}(z)$
for $\cM_\ell$ is identical to that of $\cM$ whenever $z\in
I_\ell$ and vanishes for $z\notin I_\ell$. Consequently, the
transfer matrix of $\cM_\ell$ is given by $\widehat{\bM}_\ell=
\widehat{\sU}(a_{\ell-1},a_\ell)$. This observation together with the fact
that $\widehat{\sU}(\infty,-\infty)=
\widehat{\sU}(\infty,a_{n-1})\widehat{\sU}(a_{n-1},a_{n-2})\cdots\widehat{\sU}(a_1,-\infty)$
proves the following theorem.
    \begin{itemize}
    \item[]{\em Theorem~2:} Let $\cM$ and $\cM_\ell$ with $\ell=1,2,\cdots,n$ be isotropic scattering media  as described above. In particular, $\cM_\ell$ has the same permittivity and permeability profiles as $\cM$ in its support, and the support of $\cM_{\ell}$ lies to the left of that of $\cM_{\ell+1}$ along the $z$-axis. Then the transfer matrix $\widehat{\bM}$ of $\cM$ can be expressed in terms of the transfer matrix $\widehat{\bM}_\ell$ of $\cM_\ell$ according to
        \be
        \widehat{\bM}=\widehat{\bM}_n\widehat{\bM}_{n-1}\cdots\widehat{\bM}_1.
        \label{compose-EM}
        \ee
    \end{itemize}
This theorem draws attention to the potential applications of the EM transfer matrix in developing numerical schemes for solving EM scattering problems.

It is important to realize that in contrast to the earlier EM transfer-matrix theories \cite{pendry-1994,mclean}, the composition rule (\ref{compose-EM}) does not enter the definition or construction of the transfer matrix we have introduced in Sec.~\ref{S3}. Consequently, the application of our EM transfer-matrix formalism does not require slicing of the medium (scatterer) along the $z$-direction, its discretization along the $x$- and $y$-directions (in position or momentum space), or the use of the composition property (\ref{compose-EM}). In principle, we can use our approach to describe the scattering of EM waves by an arbitrary isotropic medium that needs not have any particular symmetry. As we show in Secs.~\ref{S-point}-\ref{S7}, one can try to determine the transfer matrix by directly evaluating the terms in the Dyson series appearing in the right-hand side of (\ref{M=exp3}). This makes the applications described in these sections beyond the reach of the conventional transfer-matrix theories.

\section{Application for a point scatterer}
\label{S-point}

The scattering of electromagnetic waves by a point scatterer is of central importance for various applications and, as a result, has been extensively studied in the literature. Ref.~\cite{VCL} provides an illuminating review of the application of the standard Green's function methods to this problem. It is well-known that these methods lead to divergences which require renormalization of the coupling constant. In this section, we use the electromagnetic transfer-matrix method we have developed in the preceding sections to offer a solution of the scattering problem for a nonmagnetic point scatterer, i.e., the permittivity and permeability profile given by
    \begin{align}
    &\hat\varepsilon(\bbr)=1+\fz\,\delta(\bbr),
    &&\hat\mu(\bbr)=1,
    \label{d-ep=}
    \end{align}
where $\fz$ is a possibly complex coupling constant, and $\delta(\bbr)$ is the Dirac delta function in three dimensions; $\delta(\bbr):=\delta(x)\delta(y)\delta(z)$.

First, we note that according to (\ref{int-pic-H}),
    \be
    \widehat{\sH}(z)= e^{iz\widehat{\tilde{\bH}}_0}\delta\widehat{\tilde{\bH}}(z) e^{-iz\widehat{\tilde{\bH}}_0},
        \label{int-pic-H-2}
    \ee
where
     \begin{align}
        &\delta\widehat{\tilde{\bH}}(z):=\widehat{\tilde{\bH}}(z)-\widehat{\tilde{\bH}}_0=
        \left[\begin{array}{cc}
         \bzero & \delta\widehat{\tilde{\bL}}_1\\
        - \delta\widehat{\tilde{\bL}}_2& \bzero\end{array}\right],
        \label{delta-h}\\[3pt]
        &\delta\widehat{\tilde{\bL}}_1:=\widehat{ \tilde{\bL}}[\hat\varepsilon^{-1},\hat\mu]-\widehat{\tilde{\bL}}_0=
       \widehat{ \tilde{\bL}}[\hat\varepsilon^{-1}-1,\hat\mu-1]=
       \widehat{ \tilde{\bL}}[\eta_{\hat\varepsilon^{-1}},\eta_{\hat\mu}],\\[6pt]
        &\delta\widehat{\tilde{\bL}}_2:=\widehat{ \tilde{\bL}}[\hat\mu^{-1},\hat\varepsilon]-\widehat{\tilde{\bL}}_0=
       \widehat{ \tilde{\bL}}[\hat\mu^{-1}-1,\hat\varepsilon-1]=
       \widehat{ \tilde{\bL}}[\eta_{\hat\mu^{-1}},\eta_{\hat\varepsilon}],
        \end{align}
$\widehat{\tilde{\bL}}[\cdot,\cdot]$ is defined by (\ref{tL}), $\widehat{\tilde{\bL}}_0:=\widehat{\tilde{\bL}}[1,1]$, and for every function $f:\R^3\to\C$, we use the symbol $\eta_f$ to denote $f-1$, i.e.,
    \[\eta_f(\bbr):=f(\bbr)-1.\]
Notice also that $\big(\widehat{\tilde{\bL}}_0\bphi\big)(\vec p)={\tilde{\bL}}_0(\vec p)\bphi(\vec p)$, where 
$\bphi\in\sF_k^2$ is an arbitrary test function, and ${\tilde{\bL}}_0(\vec p)$ is the $2\times 2$ matrix given by (\ref{t-H-zero}).

For a point scatterer specified by (\ref{d-ep=}),
    \begin{align}
    &\eta_{\hat\varepsilon}(\bbr)=\fz\,\delta(\bbr)=\fz\,\delta(\vec r)\delta(z),
    &&\eta_{\hat\mu}(\bbr)=\eta_{\hat\mu^{-1}}(\bbr)=0.
    \label{d-q001}
    \end{align}
It is also not difficult to show that for every smooth test function $\xi:\R^3\to\C$,
    \[\int_{\R^3}d^3\bbr' \xi(\bbr')\:\eta_{\widehat{\varepsilon}^{-1}}(\bbr'-\bbr)=
    \int_{\R^3}d^3\bbr'\: \frac{-\fz\,\xi(\bbr')\delta(\bbr'-\bbr)}{1+\fz\,\delta(\bbr'-\bbr)}=
    -\frac{\fz\,\xi(\bbr)}{1+\fz\,\delta(\bzero)}=0.\]
Therefore, $\eta_{\widehat{\varepsilon}^{-1}}(\bbr)=0$.
In view of this relation, (\ref{L=}), and (\ref{int-pic-H-2}) -- (\ref{d-q001}),
    \begin{align}
    & \delta{\widehat{\tilde{\bL}}}_1=\bzero, \quad\quad
    \delta{\widehat{\tilde{\bL}}}_2=-ik\fz\,\delta(\bbr)\bsigma_2,\\[6pt]
    &\widehat{\sH}(z)=ik\fz\,\delta(z)\,
    \tilde\delta(i\vec\nabla_p)\bK,
    \label{d-sH=}
    \end{align}
where
    \begin{align}
    &\bsigma_2:=\left[\begin{array}{cc} 0 & -i\\ i & 0 \end{array}\right],
    &&\bK:=\left[\begin{array}{cc}
    \bzero & \bzero \\
    \bsigma_2 & \bzero\end{array}\right],
    \label{d-K=}
    \end{align}
%$\bsigma_2:=\mbox{\scriptsize$\left[\begin{array}{cc} 0 & -i\\ i & 0 \end{array}\right]$}$ is the second Pauli matrix,
and $\tilde\delta(i\vec\nabla p)$ is the operator acting in the function space $\cF_k^d$ according to
    \be
    \tilde\delta(i\vec\nabla_p)\bF(\vec p):=
    \frac{1}{4\pi^2}\int_{\sD_k} d\vec q\;\bF(\vec q).
    \label{d-tdelta=}
    \ee
Notice that the right-hand side of this equation does not involve $\vec p$, i.e., it takes the same value for all $\vec p\in\sD_k$.

Because $\bK^2=\bzero$, $\widehat{\sH}(z_1)\widehat{\sH}(z_2)=\bzero$. Therefore, the Dyson series expansion (\ref{M=exp3}) of the EM transfer matrix terminates, and we find
    \be
    \widehat{\bM}=\widehat\bI+k\fz\,\tilde\delta(i\vec\nabla_p)\bK=
    \left[\begin{array}{cc}
    \bI & \bzero \\
    k\fz\,\tilde\delta(i\vec\nabla_p)\bsigma_2 &\bI\end{array}\right].
    \label{d-M=}
    \ee

Next, we compute the reflection and transmission amplitudes $\bT^l_\pm(\vec p)$. To do this, we first use (\ref{d-M=}) to express (\ref{d-eq01}) in the form
    \be
    \bT^l_-(\vec p)=-k\fz\,\bPi_2(\vec p)\bX,
        \label{d-Tm=}
        \ee
where $\vec p\in\sD_k$ is arbitrary, and
    \be
    \bX:=\tilde\delta(i\vec\nabla_p)\bK\left[\bT^l_-(\vec p)+4\pi^2
        \bup_{\rm i} \delta(\vec p-\vec k_{\rm i})\right].
    \label{d-X=}
    \ee
It is important to note that, according to (\ref{d-tdelta=}), $\bX$ does not depend on $\vec p$.

Let us express $\bT^l_-(\vec p)$, $\bX$, and $\bup_{\rm i}$ as:
    \begin{align}
    &\bT^l_-(\vec p)=\left[\begin{array}{c}
    {\vec T}^+_-(\vec p) \\
    \vec T^-_-(\vec p) \end{array}\right],
    &&\bX=\left[\begin{array}{c}
    \vec X^+ \\
    \vec X^-\end{array}\right],
    &&
    \bup_{\rm i}=\left[\begin{array}{c}
    \vec\Upsilon_{\rm i}^+ \\
    \vec\Upsilon_{\rm i}^-\end{array}\right],
    \label{d-2-comp}
    \end{align}
where $\vec T^\pm_-(\vec p)$, $\vec X^\pm$, and $\vec\Upsilon_{\rm i}^\pm$ are
two-component column vectors. Then in view of (\ref{proj}), (\ref{d-K=}), (\ref{d-tdelta=}), and (\ref{d-Tm=}) -- (\ref{d-2-comp}), we have
    \begin{align}
    &\vec X^+=\vec 0,
    && \vec T_-^+(\vec p)=-\frac{k\,\fz}{2\varpi(\vec p)}\,\tilde{\bL}_0(\vec p)\vec X^-,
    && \vec T_-^-(\vec p)=-\frac{k\,\fz}{2}\,\vec X^-,
    \label{d-XTT=}
    \end{align}
where $\tilde{\bL}_0(\vec p)$ is given in (\ref{t-H-zero}). This reduces the calculation of $\bT^l_-(\vec p)$ to that of $\vec X^-$. To determine the latter, we substitute (\ref{d-2-comp})  and (\ref{d-XTT=}) in (\ref{d-X=}) and use the result together with (\ref{t-H-zero}) and (\ref{d-tdelta=}) to show that
    \bea
    \vec X^-&=&
    \bsigma_2\tilde\delta(i\vec\nabla_p)\left[\vec T_-^+(\vec p)+4\pi^2
    \vec\Upsilon_{\rm i}^+\delta(i\vec p-\vec k_i)\right]\nn\\
    &=&-\frac{k\,\fz}{2}\,
    \tilde\delta(i\vec\nabla_p)
    \left[\varpi(\vec p)^{-1}\bsigma_2\tilde{\bL}_0(\vec p)\vec X^-\right]+
    \bsigma_2\vec\Upsilon_{\rm i}^+\nn\\
    &=&\frac{ik^3\fz}{6\pi}\vec X^-+\bsigma_2\vec\Upsilon_{\rm i}^+.\nn
    \eea
This in turn implies
    \be
    \vec X^-=
    \frac{\bsigma_2\vec\Upsilon_{\rm i}^+}{1-i\fz k^3/6\pi}.
    \label{d-X-minis}
    \ee

In view of (\ref{t-H-zero}), (\ref{proj}),  and (\ref{d-2-comp}) -- (\ref{d-X-minis}),
    \bea
    \bT^l_-(\vec p)&=&
    -\frac{\fz\,k}{2 \varpi(\vec p)(1-i\fz k^3/6\pi)}
    \left[\begin{array}{c}
   \widehat{ \tilde{\bL}}_0(\vec p)\bsigma_2\vec\Upsilon^+\\
    \varpi(\vec p)\bsigma_2\vec\Upsilon^+\end{array}\right]\nn\\
    &=&-\frac{\fz\, k \bPi_2(\vec p)\bK\bup_{\rm i}}{1-i\fz k^3/6\pi}.
    \label{d-T-minus=}
    \eea
Similarly, we can use (\ref{d-eq02}) to show that
    \be
    \bT^l_+(\vec p)=k\fz\,\bPi_1(\vec p)\bX
    =\frac{\fz\, k \bPi_1(\vec p)\bK\bup_{\rm i}}{1-i\fz k^3/6\pi}.
    \label{d-T-plus=}
    \ee

Having calculated $\bT_\pm^l(\vec p)$, we can employ (\ref{f=2}) and (\ref{cross-sec=1}) to determine the scattering amplitude and differential cross section for the point scatterer. To derive a more explicit expression for these, we first establish the identities:
    \bea
    &&\varpi(\vec p)\bPi_j(\vec p)\bK=\frac{1}{2}\left[\begin{array}{cc}
    (-1)^{j+1}ik\bJ(\vec p) & \bzero\\
    \varpi(\vec p)\bsigma_2 & \bzero\end{array}\right],
    \label{id-129}\\
    &&\varpi(\vec p)^2[\bPi_j(\vec p)\bK]^\dagger\bPi_j(\vec p)\bK=
    \frac{k^2+\varpi(\vec p)^2}{4}\left[\begin{array}{cc}
    \bJ(\vec p) & \bzero\\
    \bzero & \bzero\end{array}\right],
    \label{id-130}
    \eea
where $j=1,2$ and
    \be
    \bJ(\vec p):=\frac{i}{k}\tilde{\bL}_0(\vec p)\bsigma_2=
    \frac{1}{k^2}
    \left[\begin{array}{cc}
    k^2-p_x^2 & -p_xp_y\\
    -p_xp_y & k^2-p_y^2
    \end{array}\right].
    \nn%\label{J=}
    \ee
In particular,
    \be
    \bJ(\vec k_{\rm s})=
    \left[\begin{array}{cc}
    1-\sin^2\vartheta\cos^2\varphi & -\sin^2\vartheta\sin\varphi\cos\varphi\\
    -\sin^2\vartheta\sin\varphi\cos\varphi & 1-\sin^2\vartheta\sin^2\varphi
    \end{array}\right].
    \label{J=2}
    \ee
Substituting (\ref{d-T-minus=}) and (\ref{d-T-plus=}) in
(\ref{fe=1}) and (\ref{cross-sec=1}) and making use of
(\ref{Upsilon-i=}), (\ref{d-bXi=}), and
(\ref{id-129}) -- (\ref{J=2}) we find
    \bea
%    f(\bk_{\rm i},\bk_{\rm s})\hat\bbe_{\rm s}&=&
%    \frac{k^2\left[\vec{e}_{\rm i}
%    -(\hat\bbr\cdot \vec{e}_{\rm i})\hat\bbr\right]}{4\pi\left(\fz^{-1}-i k^3/6\pi\right)}=
%    \frac{k^2\,\hat\bbr\times\left(\vec{e}_{\rm i}
%    \times\hat\bbr\right)}{4\pi\left(\fz^{-1}-i k^3/6\pi\right)},
%    \label{d-fes=}\\
     f(\bk_{\rm i},\bk_{\rm s})\hat\bbe_{\rm s}&=&
%    \frac{k^2\left[\vec{e}_{\rm i}
%    -(\hat\bbr\cdot \vec{e}_{\rm i})\hat\bbr\right]}{4\pi\left(\fz^{-1}-i k^3/6\pi\right)}=
      \frac{\ft(k)\,\left[(\hat\bbr\cdot \vec{e}_{\rm i})\hat\bbr-
      \vec{e}_{\rm i}\right]}{4\pi}=
    \frac{\ft(k)\,\hat\bbr\times\left(\hat\bbr\times\vec{e}_{\rm i}\right)}{4\pi},
    \label{d-fes=}\\
%   f(\bk_{\rm i},\bk_{\rm s})&=&\frac{\fz k^2
%    \left[\vec\Upsilon_{\rm s}^{+\dagger}\bJ(\vec k_{\rm s})-i\cos\vartheta
%    \vec\Upsilon_{\rm s}^{-\dagger}\bsigma_2\right]
%    \vec\Upsilon_{\rm i}^+}{
%    4\pi\left(1-i\fz k^3/6\pi\right)(1+\cos^2\vartheta)},
%    \label{d-f=}\\
%    \sigma_d(\bk_{\rm i},\bk_{\rm s})&=&
%    \frac{|\fz|^2 k^4\vec\Upsilon_{\rm i}^{+\dagger}\bJ(\vec k_{\rm s})\vec\Upsilon_{\rm i}^+}{16\pi^2\left|1-i\fz k^3/6\pi\right|^2},
      \sigma_d(\bk_{\rm i},\bk_{\rm s})&=&
    %\frac{k^4\left(|\vec{e}_{\rm i}|^2
    %-|\hat\bbr\cdot\vec{e}_{\rm i}|^2\right)}{16\pi^2\left|\fz^{-1}-i k^3/6\pi\right|^2}=
    \frac{|\ft(k)|^2\left(|\vec{e}_{\rm i}|^2
    -|\hat\bbr\cdot\vec{e}_{\rm i}|^2\right)}{16\pi^2},
    \label{d-cross-sec=}
    \eea
where $\vec{e}_{\rm i}$ is the projection of $\hat\bbe_{\rm
    i}$ onto the $x$-$y$ plane, i.e.,
    $\vec{e}_{\rm i}:=(\hat\bbe_x\cdot\hat\bbe_{\rm i})
    \hat\bbe_x+(\hat\bbe_y\cdot\hat\bbe_{\rm i})\hat\bbe_y=\hat\bbe_{\rm
    i}-(\hat\bbe_z\cdot\hat\bbe_{\rm i})\hat\bbe_z$,
and
    \be
    \ft(k):=\frac{-k^2}{\fz^{-1}-i k^3/6\pi}.
    \label{d-t=}
    \ee
%and for ${\rm a}={\rm i},{\rm s}$ we have introduced
%    \begin{align*}
%    \vec\Upsilon_{\rm a}^+:=\left[\begin{array}{c}
%    \hat\bbe_x\cdot\hat\bbe_{\rm a}\\
%    \hat\bbe_y\cdot\hat\bbe_{\rm a}\end{array}\right],
%    &&\vec\Upsilon_{\rm a}^-:=\left[\begin{array}{c}
%    (\hat\bbe_x\times\hat\bbr)\cdot\hat\bbe_{\rm a}\\
%    (\hat\bbe_y\times\hat\bbr)\cdot\hat\bbe_{\rm a}\end{array}\right].
 %   \end{align*}
% For an incident wave that is linearly polarized along the $x$-axis, $\hat\bbe_{\rm i}=\hat\bbe_x$, and (\ref{d-f=}) and (\ref{d-cross-sec=}) give
%    \bea
%    f(\bk_{\rm i},\bk_{\rm s})&=&
%    \frac{\fz k^2 \hat\bbe_x\cdot\hat\bbe_{\rm s}}{
%    4\pi\left(1-i\fz k^3/6\pi\right)},   \\
%    \sigma_d(\bk_{\rm i},\bk_{\rm s})&=&
%    \frac{|\fz|^2 k^4(1-\sin^2\vartheta\cos^2\varphi)}{16\pi^2\left|1-i\fz k^3/6\pi\right|^2}.
%    \eea

As we mentioned above the standard treatment of the scattering problem for the point scatterer yields a Born series involving divergent terms. Ref.~\cite{VCL} outlines a regularization of these divergences. It involves identifying the coupling constant $\fz$ of (\ref{d-ep=}) with a bare coupling constant $\fz_B$ and introducing a pair of momentum cutoffs $\Lambda_L$ and $\Lambda_T$ associated with the divergences arising from the longitudinal and transverse Green's functions. These enter in the expression for the scattered field after one sums the regularized Born series. This procedure leads to the very same formulas we have found for the scattering amplitude and differential cross-section, namely (\ref{d-fes=}) and (\ref{d-cross-sec=}), provided that we set
    \be
    \ft(k):=
    \frac{-k^2}{\fz_B^{-1}+(\Lambda_L^3-k^2\Lambda_T
    -i k^3)/6\pi}.
    \label{d-t-k-renorm}
    \ee
Comparing (\ref{d-t=}) and (\ref{d-t-k-renorm}), we see that our method is equivalent to identifying $\fz$ with the renormalized coupling constant defined by
    \be
    \fz:=\frac{\fz_B}{1+ (\Lambda_L^3-k^2\Lambda_T)\fz_B/6\pi}.
    \ee
Note, however, that we did not need to deal with any divergent terms throughout our calculations. Nor did we sum an infinite Born series after renormalizing its terms. This is an important advantage of our method over the standard Green's function approaches.

\section{Perfect broadband invisibility}
\label{S6}

A medium $\cM$ does not scatter an incident EM wave with polarization vector $\hat\bbe_{\rm i}$ and wave vector $\bk_{\rm i}$ if and only if the corresponding scattering amplitude vanishes for every choice of the polarization and wavevector of the scattered wave, $\hat\bbe_{\rm s}$ and $\hat\bk_{\rm s}$. If this happens for a finite or infinite interval of values of the wavenumber $k$ and irrespectively of the choice of $\hat\bbe_{\rm i}$ and $\hat\bk_{\rm i}$, we say that $\cM$ displays {\em broadband invisibility}. In this section, we derive a simple criterion for invisibility of an isotropic medium for wavenumbers not exceeding a prescribed critical value $\alpha$. To emphasize that our derivation does not rely on any approximation scheme, we call this phenomenon: ``perfect broadband invisibility.''

The search for broadband invisibility has a long history. Recent results on the use of conformal mappings \cite{leonhardt}, metamaterials \cite{pendry-2006,schuring-2008}, and transformation optics \cite{rahm} have led to some important progress in the subject. Our route to perfect broadband invisibility is fundamentally different from these, because as we explain below it makes use of ordinary isotropic media without invoking geometric optics arguments.

According to (\ref{f=2}) $\cM$ is invisible if and only if $\bT^{l/r}_\pm=0$. We also recall that $\bT^{l/r}_\pm$ belong to the function space $\sF^4_k$, and the EM transfer matrix $\widehat{\bM}$ is a linear operator acting in this space.

In view of (\ref{Tm=3}) and (\ref{Tp=3}), the invisibility condition,
    \be
    \bT^{l/r}_\pm=0,
    \label{invisible-condi-z23}
    \ee
holds, if $\widehat{\bM}=\widehat\bI$. Because $\widehat{\bM}$ is the time-ordered exponential of the interaction-picture Hamiltonian $\widehat{\sH}(z)$, we can satisfy (\ref{invisible-condi-z23}) by demanding that $\widehat{\sH}(z)$ vanishes identically on $\sF^4_k$. We can use (\ref{tH=1}) and (\ref{int-pic-H-2}) to express this condition in the form
    \be
    \delta\widehat{\tilde{\bH}}(z)\bF=\bzero~~~{\rm for}~~~\bF\in \sF^4_k.
    \label{condi-inv1}
    \ee
According to (\ref{delta-h}), we can fulfill (\ref{condi-inv1}), if $\delta\widehat{\tilde{\bL}}_\ell\bphi(\vec p)=\bzero$ for every $\bphi\in\cF^2_k$ and $\ell=1,2$. This in turn means that the entries $[\delta\widehat{\tilde L}_\ell]_{ij}$ of $\delta\widehat{\tilde{\bL}}_\ell$ satisfy
    \be
    [\delta\widehat{\tilde L}_\ell]_{ij}\phi(\vec p)=0~~~{\rm for}~~~\phi\in\cF^1_k.
    \label{condi-inv2}
    \ee
With the help of (\ref{L=}) and (\ref{tL}) -- (\ref{symbol}), we can
establish (\ref{condi-inv2}) by demanding that the following
requirement holds for $f=\hat\varepsilon,\hat\varepsilon^{-1},
\hat\mu$, and $\hat\mu^{-1}$.
    \be
    \int_{\sD_k}d^2\vec q\: \tilde\eta_f(\vec p-\vec q,z)\,\phi(\vec q)=0
    ~~~{\rm for}~~~\phi\in\cF^1_k.
    \label{condi-inv3}
    \ee
Making  the change of variable: $\vec q\to\vec q^{\:\prime}:=\vec p-\vec q$, we can express (\ref{condi-inv3}) as
    \be
    \int_{\sD'_k(\vec p)}d^2\vec q^{\:\prime}\: \tilde\eta_f(\vec q^{\:\prime},z)\,
    \phi(\vec p-\vec q^{\:\prime})=0
    ~~~{\rm for}~~~\phi\in\cF^1_k,
    \label{condi-inv4}
    \ee
where $\sD'_k(\vec p):=\big\{\vec q^{\:\prime}\in\R^2~\big|~|\vec q^{\:\prime}-\vec p|<k\:\big\}$.

In summary, (\ref{condi-inv4}) is a sufficient condition for the vanishing of the scattering amplitude. The following invisibility theorem is a direct consequence of this condition.
    \begin{itemize}
    \item[]{\em Theorem~3:}  Consider an isotropic scattering medium  $\cM$ with relative permittivity and permeability profiles $\hat\varepsilon$ and $\hat\mu$. Let $\eta_{f}:=f-1$, and $\alpha$ be a given wavenumber scale. Suppose that for $f=\hat\varepsilon,\hat\varepsilon^{-1}, \hat\mu$, and $\hat\mu^{-1}$, the Fourier transform of $\eta_f(\vec r,z)$ with respect to $\vec r$, which we denote by $\tilde \eta_f(\vec\fK,z)$,
satisfies:
    \be
    \tilde \eta_f(\vec\fK,z)=0~~~{\rm for}~~~|\vec\fK|< 2\alpha.
    \label{condi-inv5}
    \ee
Then $\cM$ does not scatter any incident EM plane wave whose wavenumber $k\leq\alpha$.
    \item[]{\em Proof:} Suppose that $k\leq\alpha$. Then for all $\vec p\in\sD_k$ and $\vec q^{\:\prime}\in\sD'_k(\vec p)$, we have $\vec q:=\vec p-\vec q^{\:\prime}\in\sD_k$ and
$|\vec q^{\:\prime}|=|\vec p-\vec q|\leq|\vec p|+|\vec q|< 2k\leq
2\alpha$. This relation together with the hypothesis of the theorem
imply that the $\tilde\eta_f(\vec q^{\,\prime},z)$ appearing in
(\ref{condi-inv4}) vanishes for
$f=\hat\varepsilon,\hat\varepsilon^{-1}, \hat\mu$, and
$\hat\mu^{-1}$. Therefore (\ref{condi-inv4}) holds, and $\cM$ is
invisible for incident plane waves with $k\leq\alpha$.~~~$\square$
    \end{itemize}
The characterization of functions $f(\vec r,z)$ with $\tilde f(\vec\fK,z)=\widetilde{f^{-1}}(\vec\fK,z)=0$ for $|\vec\fK|\leq 2\alpha$ is not easy. In the following we present a slightly weaker invisibility theorem that allows for a simple construction of permittivity and permeability profiles displaying perfect broadband invisibility. We give a proof of this theorem in Appendix~C.
    \begin{itemize}
    \item[]{\em Theorem~4:} Let $\cM$, $\hat\varepsilon$, $\hat\mu$, $\eta_{\varepsilon}$, and $\eta_{\mu}$ be as in Theorem~3, $\alpha_x$ and $\alpha_y$ be a pair of wavenumber scales, $\alpha$ be the smallest of $\alpha_x$ and $\alpha_y$, and $\tilde f(\fK_x,y,z)$ and $\tilde f(x,\fK_y,z)$ denote the Fourier transform of $f(x,y,z)$ with respect to $x$ and $y$, respectively. Suppose that $\hat\varepsilon$ and $\hat\mu$ are bounded functions whose real part has a positive lower bound, and the following conditions hold for $f=\hat\varepsilon$ and $\hat\mu$.
        \be
        \begin{aligned}
        &\tilde\eta_f(\fK_x,y,z)=0~~~{\rm for}~~~\fK_x< 2\alpha_x,\\
        &\tilde\eta_f(x,\fK_y,z)=0~~~{\rm for}~~~\fK_y< 2\alpha_y.
        \end{aligned}
        \label{condi-inv6}
        \ee
Then $\cM$ does not scatter any incident EM plane wave whose wavenumber $k\leq\alpha$.
    \end{itemize}

It is easy to check that (\ref{condi-inv6}) is equivalent to
    \be
    \begin{aligned}
    &\hat\varepsilon(x,y,z)=e^{2i\alpha_x x}u_\varepsilon(x,y,z)+
    e^{2i\alpha_y y}v_\varepsilon(x,y,z)+1,\\
    &\hat\mu(x,y,z)=e^{2i\alpha_x x}u_\mu(x,y,z)+
    e^{2i\alpha_y y}v_\mu(x,y,z)+1,
    \end{aligned}
    \label{u-v}
    \ee
where $u_\varepsilon,u_\mu,v_\varepsilon,v_\mu:\R^3\to\C$ are functions fulfilling:
    \be
    \begin{aligned}
    &\tilde u_f(\fK_x,y,z)=\tilde v_f(\fK_x,y,z)=0~~~{\rm for}~~~\fK_x<0,\\
    &\tilde u_f(x,\fK_y,z)=\tilde v_f(x,\fK_y,z)=0~~~{\rm for}~~~\fK_y<0.
    \end{aligned}
    \label{condi-inv7}
    \ee
We can construct concrete examples of such functions by noting that they are inverse Fourier transform of functions $\tilde w(\fK_x,\fK_y,z)$ vanishing for $\fK_x<0$ and $\fK_y<0$, i.e., they have the generic form:
    \be
    w(\vec r,z)=\frac{1}{4\pi^2}\int_0^\infty d\fK_x\int_0^\infty d\fK_y\,
    e^{i\vec\fK\cdot\vec r}\tilde w(\fK_a,\fK_y,z),
    \label{tw=}
    \ee
where $\tilde w:\R^3\to\C$ is any function such that $\int_0^\infty d^2\vec\fK\,|\tilde w(\vec\fK,z)|<\infty$. A typical example is
    \begin{align}
    &\tilde w(\fK_x,\fK_y,z)=\tilde\fz\, e^{-\vec a\cdot\vec\fK}\fK_x^{n_x}\fK_x^{n_y}
    \chi_{a_z}(z),
    &&\chi_{a_z}(z):=\left\{\begin{array}{cc}
    1 & {\rm for}~z\in[0,a_z],\\
    0 &{\rm otherwise},\end{array}\right.
    \label{w=}
    \end{align}
where $\bba=(\vec a,a_z)\in\R^3$, $\tilde\fz\in\C$, and $n_x$ and $n_y$ are positive integers. Substituting (\ref{w=}) in (\ref{tw=}), we find
    \be
    w(\vec r,z)=\frac{\fz\,\chi_{a_z}(z)}{
    \left(x/a_x+i\right)^{n_x+1}
    \left(y/a_y+i\right)^{n_y+1}},
    \label{w=2}
    \ee
where $\fz:=n_x!n_y!\,\tilde\fz/[4\pi^2(-ia_x)^{n_x+1}(-ia_y)^{n_y+1}]$. Our analysis shows that if the relative permittivity and permeability of an isotropic medium $\cM$ is given by (\ref{u-v}), and
$u_\varepsilon,u_\mu,v_\varepsilon$, and $v_\mu$ have the form (\ref{w=2}) with possibly different choices for $\fz$, $\bba$, $n_x$, and $n_y$, then $\cM$ will be invisible for every incident plane wave whose wavenumber $k$ is smaller than both $\alpha_x$ and $\alpha_y$. It is not difficult to see that $\cM$ describes a slab of thickness $a_z$ that occupies the space between the planes $z=0$ and $z=a_z$ and is surrounded by vacuum.

The above characterization of broadband invisibility in isotropic media generalizes the results of \cite{ol-2017} on the construction of effectively two-dimensional isotropic media that are invisible for the TE and TM waves with wavenumber not exceeding a critical value. This construction also involves functions whose Fourier transform
vanishes on the negative real axis. The relevance of these functions to fullband invisibility in effectively one-dimensional optical systems has been originally noted in \cite{horsley}. See also
\cite{longhi1,longhi2,longhi3,hl-review}. Another notable aspect of our construction is that it does not rely on any approximation scheme; the broadband invisibility displayed by these media is absolutely exact.

\section{$\boldsymbol{\alpha}$-Equivalent Scattering Media}
\label{S7}

In the preceding section we have obtained simple criteria for the broadband invisibility of isotropic media for wavenumbers $k\leq\alpha$, where $\alpha$ is an arbitrary preassigned wavenumber scale. As far as its EM scattering properties are concerned such a medium is equivalent to vacuum whenever the incident wave is an EM plane wave with wavenumber not exceeding $\alpha$ or a superposition of such plane waves. In the following we extend this notion of equivalence to a pair of scattering media.
    \begin{itemize}
    \item[]{\em Definition~2:} Let $\alpha$ be a wavelength scale, and $\cM_1$ and $\cM_2$ be a pair of isotropic scattering media. $\cM_1$ and $\cM_2$ are said to be {\em $\alpha$-equivalent} if they
have the same scattering amplitude for every incident plane wave whose wavenumber does not exceed $\alpha$.
    \end{itemize}
The transfer-matrix formulation of the scattering of EM waves provides a simple characterization of the $\alpha$-equivalence of scattering media. This is the electromagnetic generalization of the notion of $\alpha$-equivalent potentials we have recently developed in Ref.~\cite{jmp-2019}.

Let us label the relative permittivity and permeability of the
medium $\cM_\ell$ by $\hat\varepsilon_\ell$ and $\hat\mu_\ell$,
respectively, where $\ell=1,2$. If $\cM_1$ and $\cM_2$ have the same
transfer matrix for $k\leq\alpha$, then they are
$\alpha$-equivalent. Because the transfer matrix is the S-matrix for
a corresponding effective quantum system, the requirement that the
Hamiltonian operators $\widehat{\tilde{\bH}}_1(z)$ and
$\widehat{\tilde{\bH}}_2(z)$ for $\cM_1$ and $\cM_2$ coincide
ensures the equality of their transfer matrices and hence their
$\alpha$-equivalence. In view of (\ref{tH=1}), we can ensure
$\widehat{\tilde{\bH}}_1(z)=\widehat{\tilde{\bH}}_2(z)$ for a given
$k$ by demanding that for every $\bxi\in\sF_k^2$,
    \begin{align*}
    &\widehat{\tilde{\bL}}[\hat\varepsilon_1^{-1},\hat\mu_1]\bxi(\vec p)=\widehat{\tilde{\bL}}[\hat\varepsilon_2^{-1},\hat\mu_2]\bxi(\vec p),
    &&\widehat{\tilde{\bL}}[\hat\mu_1^{-1},\hat\varepsilon_1]\bxi(\vec p)=\widehat{\tilde{\bL}}[\hat\mu_2^{-1},\hat\varepsilon_2]\bxi(\vec p).
    \end{align*}
Because, according to (\ref{L=}),  ${\widehat{\bL}}[f,g]$ is a linear function of both $f$ and $g$, these relations are equivalent to
    \begin{align}
    &\widehat{\tilde{\bL}}[\hat\varepsilon_1^{-1}-\hat\varepsilon_2^{-1},\hat\mu_1-\hat\mu_2]
    \bxi(\vec p)=\bzero,
    &&\widehat{\tilde{\bL}}[\hat\mu_1^{-1}-\hat\mu_2^{-1},\hat\varepsilon_1-\hat\varepsilon_2]
    \bxi(\vec p)=\bzero.
    \label{condi1023}
    \end{align}
As we explain in our proof of Theorem~3, we can satisfy
(\ref{condi1023}) for $k\leq\alpha$ provided that
$\tilde\eta(\vec\fK,z)=0$ for $|\vec\fK|\leq 2\alpha$ and
$\eta=\hat\varepsilon_1-\hat\varepsilon_2,~ \hat\mu_1-\hat\mu_2,
~\hat\varepsilon_1^{-1}-\hat\varepsilon_2^{-1}$, and
$\hat\mu_1^{-1}-\hat\mu_2^{-1}$. This proves the following theorem.
    \begin{itemize}
    \item[]{\em Theorem~5:} Let $\cM_1$ and $\cM_2$ be scattering media with $\hat\varepsilon_\ell$ and
$\hat\mu_\ell$ respectively denoting the relative permittivity and
permeability of $\cM_\ell$. Then $\cM_1$ and $\cM_2$ are
$\alpha$-equivalent, if the following condition holds for
$\eta=\hat\varepsilon_1-\hat\varepsilon_2,~
\hat\mu_1-\hat\mu_2,~\hat\varepsilon_1^{-1}-\hat\varepsilon_2^{-1}$,
and $\hat\mu_1^{-1}-\hat\mu_2^{-1}$.
    \be
    \tilde\eta(\vec\fK,z)=0~~~{\rm for}~~~|\vec\fK|\leq 2\alpha.
    \label{condi-equiv1}
    \ee
    \end{itemize}
Combining the content of Theorems~4 and 5 we are led to the following theorem on the construction of $\alpha$-equivalent    pairs of scattering media.
    \begin{itemize}
    \item[]{\em Theorem~6:} Let $\cM_\ell$ with $\ell=1,2,3,4$ be isotropic scattering media with relative permittivity and permeability, $\hat\varepsilon_\ell$ and
$\hat\mu_\ell$. Suppose that $\cM_{3}$ and $\cM_{4}$ satisfy the
hypothesis of Theorem~4, i.e., (\ref{condi-inv6}) holds for
$f=\hat\varepsilon_3,\hat\varepsilon_4,\hat\mu_3$, and $\hat\mu_4$,
and
    \begin{align}
    &\hat\varepsilon_1=\frac{\eta_{\varepsilon_3}}{2}\,\left(\sqrt{1+4/\eta_{\varepsilon_3}\eta_{\varepsilon_4}}+1\right),
    &&\hat\varepsilon_2=\frac{\eta_{\varepsilon_3}}{2}\,\left(\sqrt{1+4/\eta_{\varepsilon_3}\eta_{\varepsilon_4}}-1\right),
    \label{thm6-1}\\
    &\hat\mu_1=\frac{\eta_{\mu_3}}{2}\,\left(\sqrt{1+4/\eta_{\mu_3}\eta_{\mu_4}}+1\right),
    &&\hat\mu_2=\frac{\eta_{\mu_3}}{2}\,\left(\sqrt{1+4/\eta_{\mu_3}\eta_{\mu_4}}-1\right), \label{thm6-2}
    \end{align}
where $\eta_f:=f-1$.
Then $\cM_1$ and $\cM_2$ are $\alpha$-equivalent.
    \item[]{\em Proof:} We know that (\ref{condi-inv6}) and consequently (\ref{condi-inv5}) hold for
    $f=\hat\varepsilon_{3,4}$ and $\hat\mu_{3,4}$. We use this observation to set  $\hat\varepsilon_1-\hat\varepsilon_2=\eta_{\varepsilon_3}$ and $\hat\varepsilon_1^{-1}-\hat\varepsilon_2^{-1}=-\eta_{\varepsilon_4}$. Solving these equations for $\hat\varepsilon_1$ and $\hat\varepsilon_2$, and demanding that $\RE(\hat\varepsilon_\ell)>0$ yield (\ref{thm6-1}). Replacing the role of  $\hat\varepsilon_\ell$ in this argument  by $\hat\mu_\ell$, we obtain (\ref{thm6-2}). By construction this choice for $\hat\varepsilon_{1,2}$ and $\hat\mu_{1,2}$ fulfills the hypothesis of Theorem~5. Therefore $\cM_1$ and $\cM_2$ are $\alpha$-equivalent.~~~$\square$
\end{itemize}

\section{Concluding Remarks}
\label{S8}

Transfer matrices are used in the potential scattering in one
dimension mainly because of their composition property. A recent
study of the similarity between this property and the composition
rule for evolution operators in quantum mechanics has led to the
identification of the transfer matrix with the S-matrix of a
nonunitary effective two-level system \cite{pra-2014a,ap-2014}. This
curious fact has in turn paved the way for the introduction of a
multidimensional generalization of the transfer matrix and a
corresponding transfer-matrix formulation of potential scattering in
two and three dimensions \cite{pra-2016}. The latter is a genuine
alternative to the standard (Lipman-Schwinger) approach to scattering theory with many interesting applications \cite{prsa-2016,pra-2017,jpa-2018,ol-2017,jmp-2019}.

In the present article we have developed a transfer-matrix
formulation of the scattering of EM waves by general isotropic
media. This is a highly nontrivial generalization of the
transfer-matrix approach to the scattering of scalar waves developed in Ref.~\cite{pra-2016}. We have shown that this EM transfer matrix shares the
basic features of its analog for the scalar waves. In particular, it has a
similar composition property which should facilitate its numerical
implementations. Note however that this property and the related
slicing of the scattering medium is by no means essential for the
application of this approach in dealing with specific EM scattering
problems. The latter involves addressing two basic problems:
    \begin{enumerate}
    \item Determining the transfer matrix $\widehat{\bM}$ which is in general
an integral operator acting in the function space $\sF_k^4$. This
happens to coincide with the S-matrix for an effective non-unitary
quantum system and admits a Dyson series expansion of the form
(\ref{M=exp3}).
    \item Solving the integral equations (\ref{Tm=2}) and (\ref{Tm=2r})
and substituting the result in (\ref{Tp=2}) and (\ref{Tp=2r}) to determine
$\bT_{\pm}^{l/r}$, which in turn yield the scattering amplitude for the
left- and right-incident waves.
    \end{enumerate}
Developing various exact, approximate, and numerical methods of
achieving these will play an important role in making the EM transfer-matrix
formalism into a mainstream method of solving EM scattering problems.

A most interesting application of the transfer matrix of Ref.~\cite{pra-2016} is that it leads to an exact
solution of the scattering problem for the $\delta$-function potentials in two and three dimensions while avoiding the divergences of the conventional approach
\cite{pra-2016,jpa-2018}. In a sense, this transfer matrix
method has a built-in regularization mechanism. The EM transfer
matrix we have introduced in the present article has the same
property; it yields an exact and finite expression for the
scattering amplitude of a non-magnetic delta-function point scatterer
which agrees with the known results after we identify the original coupling constant of our approach with a renormalized coupling constant of the Green's function methods. This reveals a striking advantage of our method, because its application to a point scatterer does not require dealing with divergent quantities.

Another concrete evidence for the effectiveness of our approach to EM scattering is its role in the discovery of a large class of isotropic media that display perfect broadband invisibility for wavenumbers $k$ not exceeding a prescribed value $\alpha$. For
$k\leq\alpha$ such a medium behaves exactly like vacuum. Motivated
by this observation, we have introduced the notion of
$\alpha$-equivalent media, which share the same scattering features
for wavenumbers $k\leq\alpha$. We have employed our transfer-matrix
formulation of the scattering of EM waves to obtain a simple
quantitative scheme for constructing $\alpha$-equivalent media.

The utility of our EM transfer matrix theory in dealing with basic problems such as the singularity-free treatment of point scatterers, the characterization of isotropic media displaying exact broadband invisibility, and the study of $\alpha$-equivalent media reveals some of its advantages over the previously studied transfer matrix theories. Because the application of the latter for a general inhomogeneous medium requires slicing the medium, discretization of the slices, and the numerical evaluation of their transfer matrices, these theories cannot be effectively used for performing exact and analytic calculations. This in turn limits their effectiveness in dealing with the type of basic problems we address in Secs.~\ref{S-point}-\ref{S7}.
 \vspace{6pt}

\section*{Appendix~A: Direct calculation of differential cross section}

In this appendix, we offer an alternative derivation of Eq.~(\ref{cross-sec=1}) that avoids the calculation of the scattering amplitude. First, we note that according to (\ref{diff-cross-sec}), (\ref{Es=4}), and (\ref{Hs=4}),
    \begin{align}
    &|\bcE_{\rm s}(\bbr)|^2+|\bcH_{\rm s}(\bbr)|^2=\frac{2|\cE_0|^2}{r^2}\,\sigma_d(\bk_{\rm i},\bk_{\rm s})
    ~~~~~{\rm for}~~~~r\to\infty.
    \label{Abs-f2}
    \end{align}
Because $\{\hat\bbr,\hat\bbe_{\rm s},\hat\bbr\times\hat\bbe_{\rm s}\}$ forms an orthonormal basis of $\R^3$, Pythagorean theorem states that
    \be
    (\hat\bbe_z\cdot\hat\bbe_{\rm s})^2+[\hat\bbe_z\cdot(\hat\bbr\times\hat\bbe_{\rm s})]^2=
    1-(\hat\bbe_z\cdot\hat\bbr)^2=1-\cos^2\vartheta.
    \label{id-31}
    \ee
Equations~(\ref{Ei-Es}), (\ref{bPhi-s}), (\ref{Hs=4}), (\ref{Abs-f2}), and (\ref{id-31}) allow us to relate $|f(\bk_{\rm i},\bk_{\rm s})|^2$ to the asymptotic expression for $|\bPhi_{\rm s}(\bbr)|^2:=\bPhi_{\rm s}(\bbr)^\dagger\bPhi_{\rm s}(\bbr)$. The result is:
    \bea
    |\bPhi_{\rm s}(\bbr)|^2
    &=&
    |\vec\cE_{\rm s}(\bbr)|^2+|\vec\cH_{\rm s}(\bbr)|^2
    %,\nn\\
    %&=&
    =|\bcE_{\rm s}(\bbr)|^2+|\bcH_{\rm s}(\bbr)|^2-
    |\hat\bbe_z\cdot\bcE_{\rm s}(\bbr)|^2-
    |\hat\bbe_z\cdot\bcH_{\rm s}(\bbr)|^2\nn\\
    &=&\frac{|\cE_0|^2}{r^2}\,(1+\cos^2\vartheta)\,
    \sigma_d(\bk_{\rm i},\bk_{\rm s})
    ~~~~~{\rm for}~~~~r\to\infty.
    \eea
Substituting (\ref{bPhi-s2}) in this equation and solving for $\sigma_d(\bk_{\rm i},\bk_{\rm s})$, we find (\ref{cross-sec=1}).

\section*{Appendix~B: Integral kernels for $\widehat{\tilde{\bH}}(z)$ and $\widehat{\sH}(z)$}

The Hamiltonian operators $\widehat{\tilde{\bH}}(z)$ and $\widehat{\sH}(z)$ are integral operators acting on the function space $\sF^4_k$. We can express them in terms of the corresponding integral kernels, ${\tilde{\bH}}(z;\vec p,\vec q)$ and ${\sH}(z;\vec p,\vec q)$. The latter determine the action of $\widehat{\tilde{\bH}}(z)$ and $\widehat{\sH}(z)$ on the four-component fields $\bF\in\sF^4_k$ according to
    \begin{align}
    &\big(\widehat{\tilde{\bH}}(z)\bF\big)(\vec p)=\int_{\sD_k}d^2\vec q\;{{\tilde{\bH}}}(z;\vec p,\vec q)\bF(\vec q),
    &&\big(\widehat{\sH}(z)\bF\big)(\vec p)=\int_{\sD_k}d^2\vec q\;{\sH}(z;\vec p,\vec q)\bF(\vec q).
    \label{kernels-appB}
    \end{align}
In this appendix we derive explicit expressions for the integral kernels ${\tilde{\bH}}(z;\vec p,\vec q)$ and ${\sH}(z;\vec p,\vec q)$.

First, we observe that according to (\ref{tH=1}), ${\tilde{\bH}}(z;\vec p,\vec q)$ has the following structure.
    \be
   {{\tilde{\bH}}}(z;\vec p,\vec q)=\left[\begin{array}{cc}
    \bzero &{{\tilde{\bh}}}_+(z;\vec p,\vec q)\\
   {{\tilde{\bh}}}_-(z;\vec p,\vec q)&\bzero\end{array}\right],
    \label{tH=1-appB1}
    \ee
where ${\tilde{\bh}}_\pm(z;\vec p,\vec q)$ are $2\times 2$ matrices depending on $z,\vec p$, and $\vec q$. In light of this equation, if we express $\bF$ in the form,
    \be
    \bF=\left[\begin{array}{c}
    \bphi_+\\
    \bphi_-\end{array}\right],
    \label{F=xi-zeta}
    \ee
where $\bphi_\pm\in\sF^2_k$, we have
    \be
  {{\tilde{\bH}}}(z;\vec p,\vec q)\bF(\vec q)=\left[\begin{array}{c}
   {{\tilde{\bh}}}_+(z;\vec p,\vec q)\bphi_-(\vec q)\\
   { {\tilde{\bh}}}_-(z;\vec p,\vec q)\bphi_+(\vec q)\end{array}\right].
    \label{tHK=-appB-1}
    \ee

Next, we introduce:
    \begin{align}
    &\widehat{\tilde\bL}_+:=\widehat{\tilde{\bL}}[\hat\varepsilon^{-1},\hat\mu],
    &&\widehat{\tilde\bL}_-:=\widehat{\tilde{\bL}}[\hat\mu^{-1},\hat\varepsilon],
    \label{cLs=appB}
    \end{align}
so that, according to (\ref{tH=1}),
    \be
    \big(\widehat{\tilde{\bH}}(z)\bF\big)(\vec p)=\left[\begin{array}{c}
    (\widehat{\tilde\bL}_+\bphi_-)(\vec p)\\
    -(\widehat{\tilde\bL}_-\bphi_+)(\vec p)\end{array}\right].
    \label{tH=1-appB}
    \ee

We can use (\ref{tL}), (\ref{integral-op1}), and (\ref{kernels-appB}) -- (\ref{tH=1-appB}) to infer that
    \be
   \left (\widehat{\tilde\bL}_\pm\bphi_\mp\right)(\vec p)=\frac{1}{4\pi^2}\int_{\sD_k}d^2\vec q\;
    \tilde\bfS_\pm(z;\vec p-\vec q,i\vec q)\bphi_\mp(\vec q),
    \label{Lpm-appB}
    \ee
where
    \bea
    \tilde\bfS_\pm(z;\vec\fK,i\vec q)&:=&\int_{\R^2}d^2\vec r\;e^{-i\vec\fK\cdot\vec r}
    \bfS_\pm(z;\vec r,i\vec q),
    \label{TfSpm=}
    \eea
and $\bfS_\pm(z;\vec r,i\vec q)$ are $2\times 2$ matrices with the entries $\fS_{\pm,ij}(z;\vec r,i\vec q)$ given by
    \bea
    \fS_{+,11}(z;\vec r,i\vec q)&:=& k^{-1}q_y\left\{-\hat\varepsilon(\vec r,z)^{-1}q_x+i\partial_x[ \hat\varepsilon(\vec r,z)^{-1}]\right\},\nn\\
    \fS_{+,12}(z;\vec r,i\vec q)&:=& k^{-1}\left\{
    \hat\varepsilon(\vec r,z)^{-1}q_x^2-iq_x\partial_x[\hat\varepsilon(\vec r,z)^{-1}]-k^2\hat\mu(\vec r,z)\right\},\nn\\
    \fS_{+,21}(z;\vec r,i\vec q)&:=& k^{-1}\left\{-\hat\varepsilon(\vec r,z)^{-1}q_y^2+iq_y\partial_y[\hat\varepsilon(\vec r,z)^{-1}]
    +k^2\hat\mu(\vec r,z)\right\},\nn\\
    \fS_{+,22}(z;\vec r,i\vec q)&:=& k^{-1}q_x\left\{\hat\varepsilon(\vec r,z)^{-1}q_y-i\partial_y[\hat\varepsilon(\vec r,z)^{-1}]\right\},\nn\\
    \fS_{-,11}(z;\vec r,i\vec q)&:=& k^{-1}q_y\left\{-\hat\mu(\vec r,z)^{-1}q_x+i\partial_x[ \hat\mu(\vec r,z)^{-1}]\right\},\nn\\
    \fS_{-,12}(z;\vec r,i\vec q)&:=& k^{-1}\left\{
    \hat\mu(\vec r,z)^{-1}q_x^2-iq_x\partial_x[\hat\mu(\vec r,z)^{-1}]-k^2\hat\varepsilon(\vec r,z)\right\},\nn\\
    \fS_{-,21}(z;\vec r,i\vec q)&:=& k^{-1}\left\{-\hat\mu(\vec r,z)^{-1}q_y^2+iq_y\partial_y[\hat\mu(\vec r,z)^{-1}]
    +k^2\hat\varepsilon(\vec r,z)\right\},\nn\\
    \fS_{-,22}(z;\vec r,i\vec q)&:=& k^{-1}q_x\left\{\hat\mu(\vec r,z)^{-1}q_y-i\partial_y[\hat\mu(\vec r,z)^{-1}]\right\}.\nn
    \eea

In view of (\ref{kernels-appB}) and (\ref{tH=1-appB}) -- (\ref{TfSpm=}), we have
    \be
    \tilde\bh_\pm(z;\vec p,\vec q)=\pm\frac{1}{4\pi^2}\,\tilde\bfS_\pm(z;\vec p-\vec q,i\vec q)=
    \pm\frac{1}{4\pi^2}\int_{\R^2}d^2\vec r\;e^{-i(\vec p-\vec q)\cdot\vec r}
    \bfS_\pm(z;\vec r,i\vec q).
    \label{kernel-bh-pm}
    \ee
Substituting this equation in (\ref{tH=1-appB1}), we find the
following expression for the integral kernel of the Hamiltonian
operator $\tilde\bH(z)$.
    \be
    \tilde\bH(z;\vec p,\vec q)=\frac{1}{4\pi^2}\int_{\R^2}d^2\vec r\;e^{-i(\vec p-\vec q)\cdot\vec r}
    \left[\begin{array}{cc}
    \bzero & \bfS_+(z;\vec r,i\vec q)\\
    -\bfS_-(z;\vec r,i\vec q)&\bzero\end{array}\right].
    \label{kernel-bH=}
    \ee

Next, we determine the integral kernel $\sH(z;\vec p,\vec q)$ for
the interaction-picture Hamiltonian $\widehat\sH(z)$. It is easy to see that
according to (\ref{H-zero-op-def}), (\ref{int-pic-H}), and
(\ref{kernels-appB}),
    \bea
    \big(\widehat\sH(z)\bF\big)(\vec p)&=&\big(e^{iz\widehat{\tilde\bH}_0}
    \widehat{\tilde\bH}(z)e^{-iz\widehat{\tilde\bH}_0}\bF\big)(\vec p)-
    \tilde\bH_0(\vec p)\bF(\vec p)\nn\\
    &=&\int_{\sD_k}d^2\vec q\: \left[e^{iz\tilde\bH_0(\vec p)}
    \tilde\bH(z;\vec p,\vec q)e^{-iz\tilde\bH_0(\vec q)}-
    \delta(\vec p-\vec q)\tilde\bH_0(\vec q)\right]\bF(\vec q),
    \label{kernel-sH=1}
    \eea
where $\tilde\bH_0(\vec p)$ is the $4\times 4$ matrix given by
(\ref{t-H-zero}), and $\delta(\vec p)$ is the Dirac delta-function in
two dimensions. In view of (\ref{kernels-appB}) and
(\ref{kernel-sH=1}),
    \be
    \sH(z;\vec p,\vec q)=e^{iz\tilde\bH_0(\vec p)}\tilde\bH(z;\vec p,\vec q)e^{-iz\tilde\bH_0(\vec q)}-
    \delta(\vec p-\vec q)\tilde\bH_0(\vec q).
    \ee

\section*{Appendix~C: Proof of Theorem~4}

To prove Theorem~4 we use the following lemmas.
    \begin{itemize}
    \item[]{\em Lemma~1:} Let $\eta_{1},\eta_{2}:\R^3\to\C$ be functions satisfying
        \begin{align}
        &\tilde\eta_{1}(\fK_x,y,z)=\tilde\eta_{2}(\fK_x,y,z)=0~~~{\rm for}~~~\fK_x< 2\alpha_x,
        \label{condi-inv6x}\\
        &\tilde\eta_{1}(x,\fK_y,z)=\tilde\eta_{2}(x,\fK_y,z)=0~~~{\rm for}~~~\fK_y< 2\alpha_y,
        \label{condi-inv6y}
        \end{align}
    and $\eta_3:=\eta_1\eta_2$. Then
        \begin{align}
        &\tilde\eta_{3}(\fK_x,y,z)=0~~~{\rm for}~~~\fK_x< 2\alpha_x,
        \label{condi-inv6x3}\\
        &\tilde\eta_{3}(x,\fK_y,z)=0~~~{\rm for}~~~\fK_y< 2\alpha_y.
        \label{condi-inv6y3}
        \end{align}
    \item[]{\em Proof:} According to (\ref{condi-inv6x}) and the convolution formula for        Fourier transform,
        \[\tilde\eta_3(\fK_x,y,z)=\int_{-\infty}^\infty dq\:\tilde\eta_1(\fK_x-q,y,z)
        \tilde\eta_2(q,y,z)=\int_{2\alpha_x}^\infty dq\:\tilde\eta_1(\fK_x-q,y,z)
        \tilde\eta_2(q,y,z).\]
    The $\fK_x-q$ appearing in the latter integral takes values in the interval
    $(-\infty,\fK_x-2\alpha)$. For $\fK_x<2\alpha_x$, $\fK_x-q<4\alpha$. Therefore,
    (\ref{condi-inv6x}) implies that $\tilde\eta_1(\fK_x-q,y,z)=0$. This proves
    (\ref{condi-inv6x3}). Swapping the roles of $x$ and $y$ in this argument yields a
    proof of (\ref{condi-inv6y3}).~~~$\square$
    \end{itemize}

    \begin{itemize}
    \item[]{\em Lemma~2:} If $f:\R\to\C$ is a bounded function whose real part
    is bounded below by a positive number, then $[1+\eta_{f}(\bbr)]^{-1}$ admits a
    convergent series expansion in powers of a degree-1 polynomial in $\eta_f(\bbr)$.
    \item[]{\rm Proof:} By the hypothesis of the lemma, there are real numbers $m$ and $M$ such that for all $\bbr\in\R^3$,
        \be
        0<m\leq\RE[f(\bbr)]\leq |f(\bbr)|\leq M.
        \label{bound}
        \ee
    Let $\beta(\bbr):=[\eta_f(\bbr)-Q]/(1+Q)$ and $Q:=(M^2+1)/2m$. Then a simple calculation shows that
        \be
        \frac{1}{1+\eta_f}=\frac{1}{(1+Q)(1+\beta)}.
        \label{exp1}
        \ee
The right-hand side of (\ref{exp1}) admits a convergent power series in $\beta$ provided that
$|\beta(\bbr)|<1$ for all $\bbr\in\R^3$. To verify this, we use $\eta_f=f-1$ and (\ref{bound}) to establish:
        \bea
        |\eta_f(\bbr)|^2&\leq&|f(\bbr)|^2+1\leq M^2+1
        %\nn\\
        %&\leq&
        \leq\frac{M^2\RE[f(\bbr)]}{m}+1=\left(2Q -\frac{1}{m}\right)\RE[f(\bbr)]+1 \nn\\
        &<&2Q\, \RE[f(\bbr)]+1=2Q\, \RE[\eta_f(\bbr)]+ 2Q+1.\nn
        \eea
    This in turn implies
        \bea
        |\beta|= \frac{|\eta_f-Q|}{1+Q} =
        \frac{\sqrt{[\RE(\eta_f)-Q]^2+\IM(\eta)^2}}{1+Q}=
        \frac{\sqrt{|\eta_f|^2 -2Q\,\RE(\eta_f)+Q^2}}{1+Q}<1,
        \label{bnd-2}
        \eea
where we have dropped the $\bbr$-dependence of $\beta$ and $\eta_f$ for brevity. In view of (\ref{bnd-2}), we can expand the right-hand side of (\ref{exp1}) as a convergent power series in powers of $\beta(\bbr)$, i.e., for all $\bbr\in\R^3$ we have
    \be
    [1+\eta(\bbr)]^{-1}=(1+Q)^{-1}\sum_{n=0}^\infty (-1)^n\beta(\bbr)^{n}.
    \label{expand}
    \ee
$\square$
    \end{itemize}

    \begin{itemize}
    \item[]{\em Proof of Theorem~4:} It is easy to see that (\ref{condi-inv6})
    implies (\ref{condi-inv5}). Because the hypothesis of the theorem assumes
    the validity of (\ref{condi-inv6}) for $f=\hat\varepsilon$ and
    $\hat\mu$, we only need to prove (\ref{condi-inv6}) for $f=\hat\varepsilon^{-1}$ and $\hat\mu^{-1}$. We can achieve this by showing that we can replace $f$ by $f^{-1}$ in (\ref{condi-inv6}). To do this, we make use of Lemmas 1 and 2. According to (\ref{expand}) of Lemma~2 and the identity $\eta_{f^{-1}}=-\eta_f/(\eta_f+1)$, we can express $\eta_{f^{-1}}$ as a convergent power series whose terms are constant multiples of $\eta_f\beta^{n}=\eta_f(a_1\eta_f+a_0)^n$. This is a linear combination of $\eta_f,\eta_f^2,\cdots,\eta_f^{n+1}$. Lemma~1 states that these satisfy (\ref{condi-inv6}). Therefore, the same holds for $\eta_f\beta^{n}$ and terms of the above-mentioned series expansion of $\eta_{f^{-1}}$. This in turn implies that the sum of the series, i.e., $\eta_{f^{-1}}$, also satisfies (\ref{condi-inv6}). Applying this argument for $f=\hat\varepsilon$ and $\hat\mu$ we see that (\ref{condi-inv6}) holds for $f=\hat\varepsilon^{-1}$ and $\hat\mu^{-1}$.~~~$\square$
    \end{itemize}

\noindent{\bf Acknowledgements.} We are indebted to Turkish Academy
of Sciences (T\"UBA) for providing the financial support for FL's
visits to Ko\c{c} University during which a major
part of the research reported here was carried out. AM has been
supported by T\"UBA's Membership Grant.

\ed